\documentclass[ reprint, superscriptaddress, amsmath, amssymb, aps, pra, twocolumn ]{revtex4-2}

\usepackage[utf8]{inputenc}
\usepackage[T1]{fontenc}

\usepackage{graphicx}

\usepackage{dcolumn}
\usepackage{bm}
\usepackage{hyperref}
\usepackage{mleftright}
\usepackage{color}

\usepackage{amsmath}
\usepackage{amssymb}

\usepackage{tikz}
\usetikzlibrary{backgrounds, fit, decorations.pathreplacing, calc}

\begin{document}

\title{Use of Faulty States in Cat-Code Error Correction}

\author{Michael Hanks}
\email{m.hanks@imperial.ac.uk}
\affiliation{Blackett Laboratory, Imperial College London, London SW7 2AZ, United Kingdom}

\author{Soovin Lee}
\affiliation{Blackett Laboratory, Imperial College London, London SW7 2AZ, United Kingdom}

\author{Nicolo Lo Piparo}
\affiliation{Okinawa Institute of Science and Technology Graduate University, 1919-1 Tancha, Onna-son, Kunigami-gun, Okinawa 904-0495, Japan}

\author{Shin Nishio}
\affiliation{Okinawa Institute of Science and Technology Graduate University, 1919-1 Tancha, Onna-son, Kunigami-gun, Okinawa 904-0495, Japan}
\affiliation{Department of Informatics, School of Multidisciplinary Sciences, Sokendai (The Graduate University for Advanced Studies), 2-1-2 Hitotsubashi, Chiyoda-ku, Tokyo 101-8430 Japan}
\affiliation{National Institute of Informatics, 2-1-2 Hitotsubashi, Chiyoda-ku, Tokyo 101-8430, Japan}

\author{William J. Munro}
\affiliation{Okinawa Institute of Science and Technology Graduate University, 1919-1 Tancha, Onna-son, Kunigami-gun, Okinawa 904-0495, Japan}
\affiliation{National Institute of Informatics, 2-1-2 Hitotsubashi, Chiyoda-ku, Tokyo 101-8430, Japan}

\author{Kae Nemoto}
\affiliation{Okinawa Institute of Science and Technology Graduate University, 1919-1 Tancha, Onna-son, Kunigami-gun, Okinawa 904-0495, Japan}
\affiliation{National Institute of Informatics, 2-1-2 Hitotsubashi, Chiyoda-ku, Tokyo 101-8430, Japan}

\author{M.S. Kim}
\affiliation{Blackett Laboratory, Imperial College London, London SW7 2AZ, United Kingdom}

\begin{abstract}
  Bosonic codes have seen a resurgence in interest for applications as varied as fault tolerant quantum architectures, quantum enhanced sensing, and entanglement distribution.
  Cat codes have been proposed as low-level elements in larger architectures, and the theory of rotationally symmetric codes more generally has been significantly expanded in the recent literature.
  The fault-tolerant preparation and maintenance of cat code states as a stand-alone quantum error correction scheme remains however limited by the need for robust state preparation and strong inter-mode interactions.
  In this work, we consider the teleportation-based correction circuit for cat code quantum error correction.
  We show that the class of acceptable ancillary states is broader than is typically acknowledged, and exploit this to propose the use of many-component ``bridge'' states which, though not themselves in the cat code space, are nonetheless capable of syndrome extraction in the regime where non-linear interactions are a limiting factor.
\end{abstract}

\maketitle

\section{Introduction}
\label{sec:introduction}

{

Most quantum-computing architectures encode information in discrete qubit codes
\cite{ryan-andersonRealizationRealTimeFaultTolerant2021a,sundaresanDemonstratingMultiroundSubsystem2023,bluvsteinLogicalQuantumProcessor2023,googlequantumaiSuppressingQuantumErrors2023},
while Bosonic modes and cavities enable tuneable coupling or readout.
More recently, harmonic modes themselves have been proposed as encodings of finite-dimensional logical spaces
\cite{naikRandomAccessQuantum2017,duckeringVirtualizedLogicalQubits2020},
which can then be entangled together in a larger discrete lattice encoding
\cite{chamberlandBuildingFaultTolerantQuantum2022,nohLowOverheadFaultTolerantQuantum2022,xuQubitOscillatorConcatenatedCodes2023a}
and naturally interfaced in modular or distributed architectures using photonic carriers
\cite{azumaQuantumRepeatersQuantum2023a,rozpedekAllphotonicGKPqubitRepeater2023,liPerformanceRotationSymmetricBosonic2024}.
Discrete qudit codes can be embedded in the phase space of one or more harmonic oscillators
\cite{braunsteinQuantumErrorCorrection1998,cochraneMacroscopicallyDistinctQuantumsuperposition1999,gottesmanEncodingQubitOscillator2001}.
Such encodings are known as bosonic codes
\cite{pantaleoniModularBosonicSubsystem2020}.
They define a discrete logical subspace within a larger continuous-variable Hilbert space and correspond in phase space to a tiling with discrete symmetry. Here we focus on single-mode Bosonic codes. A prominent example is the cat code, whose logical states are superpositions of coherent states; multi-component cat codes generalise this structure to an N-fold rotational symmetry in phase space.

}

{

For bosonic codes with discrete number-phase symmetries (including multi-component cat codes), fault tolerant measurement of the mean photon number modulo N can be achieved either by repeating a non-destructive measurement
\cite{grimsmoQuantumComputingRotationSymmetric2020},
or via error detection and post-selection
\cite{shiFaulttolerantPreparationApproximate2019}.
However, in the absence of code concatenation, reliable diagnosis of rotation errors appears to require teleportation-based error correction (“tele-correction”)
\cite{steaneEfficientFaulttolerantQuantum1999,knillFaultTolerantPostselectedQuantum2004a}.
This complicates fully fault tolerant state preparation
\cite{grimsmoQuantumComputingRotationSymmetric2020},
since tele-correction requires not only low-weight parity measurements with locally confined errors, but also high-fidelity encoded ancillary states. 
Grimsmo et al.~\cite{grimsmoQuantumComputingRotationSymmetric2020} developed a thorough framework for fault-tolerant correction and computation with rotationally symmetric bosonic codes, but the preparation of high-fidelity ancillary states for correction remains an outstanding challenge.
}

{

A range of methods have been proposed to prepare the superposed coherent states. Two-component cat states can be produced using nonlinear interactions
\cite{cochraneMacroscopicallyDistinctQuantumsuperposition1999, munroEntangledCoherentstateQubits2000,jeongEfficientQuantumComputation2002},
projective measurements
\cite{ralphQuantumComputationOptical2003,ourjoumtsevGeneratingOpticalSchrodinger2006,ourjoumtsevGenerationOpticalSchrodinger2007,takeokaConditionalGenerationArbitrary2007,lundFaultTolerantLinearOptical2008,takahashiGenerationLargeAmplitudeCoherentState2008},
and reservoir engineering
\cite{arenzGenerationTwomodeEntangled2013,everittEngineeringDissipativeChannels2014,puriEngineeringQuantumStates2017a}.
Fewer schemes prepare more general multi-component cat-code states, typically extending these ideas via non-linear interactions
\cite{yurkeGeneratingQuantumMechanical1986,shermanPreparationDetectionMacroscopic1992,garrawayGenerationDetectionNonclassical1994,leeAmplificationMulticomponentSuperpositions1994},
or interaction with an ancilla followed by projective measurements
\cite{agarwalMesoscopicSuperpositionStates2004,hastrupDeterministicGenerationFourcomponent2020,nehraAllopticalQuantumState2022}.
In general, these methods are not intrinsically robust against errors, and must be combined with active, measurement-based correction schemes
\cite{grimsmoQuantumComputingRotationSymmetric2020,glancyTransmissionOpticalCoherentstate2004,liCatCodesOptimal2017}.

}

{

In this paper, we consider the fault tolerant initialisation of the multi-component cat codes (the meaning of fault-tolerance in this context is discussed in Appendix~\ref{sec:what_is_meant_by_fault_tolerance}).
To our knowledge, the only existing proposal relies on code concatenation
\cite{grimsmoQuantumComputingRotationSymmetric2020},
requiring complex multi-mode entanglement.
Motivated by surface-code ancilla-preparation strategies that achieve resource reduction by distilling noisy operations and using early-state post-selection \cite{itogawaEvenMoreEfficient2024,gidneyMagicStateCultivation2024},
we propose an alternative to concatenation for preparing ancillary cat-code states.
Specifically, we investigate faulty state preparation based on the generalised Yurke-Stoler schemes of
Lee et al.~\cite{leeAmplificationMulticomponentSuperpositions1994},
generating states outside of the code space as a primitive in the tele-correction procedure.

}

{

Is Section~\ref{sec:the_cat_code} we introduce the cat code and discuss the role of approximate phase measurements through heterodyne detection in the tele-correction circuit.
Section~\ref{sec:many_component_cat_states} describes the method of Yurke and Stoler for generating coherent state superpositions.
It outlines how they differ from cat code states and the role we propose for them in error correction.
Section~\ref{sec:propagation_of_phases_for_state_initialisation} describes how the relative phases that distinguish these `Yurke--Stoler states' propagate through the tele-correction circuit, and how we can account for them to recover the logical state after measurement.
Section~\ref{sec:noise_and_logical_error} provides quantitative estimates for the impact of several noise processes (phase error, non-linear over-rotation, random displacements, and loss). The danger of loss during initialisation for the fault-tolerance of the circuit is discussed, and a potential avenue for circumventing this issue is introduced.
Finally, Section~\ref{sec:discussion} concludes with a discussion of the proposal, conlusions and open questions.

}

\section{The Cat Code}
\label{sec:the_cat_code}

The cat code states were first introduced with two coherent state components by Cochrane et al.
\cite{cochraneMacroscopicallyDistinctQuantumsuperposition1999}.
These states were subsequently generalised to incorporate the imaginary axis in phase space
\cite{leghtasHardwareEfficientAutonomousQuantum2013},
and to fix logical qubit information in the relative phases
\cite{mirrahimiDynamicallyProtectedCatqubits2014},
before finally being extended to superpositions with any number of coherent state components
\cite{liCatCodesOptimal2017}.
Cat codes have been used in landmark experimental demonstrations of encoded Bosonic qubits
\cite{ofekExtendingLifetimeQuantum2016,grimmStabilizationOperationKerrcat2020}.
The cat code states for a logical qubit may be defined either as a superpositon of ${2D}$ coherent states,
\begin{align}
  \vert {\mu}_{{\alpha},{D}} \rangle
  &=
  \frac{1}{2D \sqrt{\mathcal{N}_{{\mu},{\alpha},{D}}}}
  \sum^{2{D}-1}_{{k}=0}
  e^{
      - i \pi {k}
    {\mu} 
  }
  \vert e^{\frac{i\pi{k}}{{D}}} {\alpha} \rangle
  ,
\end{align}
or in terms of Fock states as
\begin{align}
  \vert {\mu}_{{\alpha},{D}} \rangle
  &=
  \frac{e^{-\frac{\lvert\alpha\rvert^{2}}{2}}}{\sqrt{\mathcal{N}_{{\mu},{\alpha},{D}}}}
  \sum^{\infty}_{{m}=0}
  \frac{
    {\alpha}^{ \left( 2 {m} + {\mu} \right) {D} }
  }{
    \sqrt{
      \left(
        \left( 2 {m} + {\mu} \right) {D}
      \right) !
    }
  }
  \vert \left( 2 {m} + {\mu} \right) {D} \rangle
  .
\end{align}
Here ${\mu}\in\{0,1\}$, the normalisation constant is
\begin{align}
  \mathcal{N}_{{\mu},{\alpha},{D}}
  &=
  e^{-\lvert\alpha\rvert^{2}}
  \sum^{\infty}_{{m}=0}
  \frac{
    {\lvert \alpha \rvert}^{ 2 \left( 2 {m} + {\mu} \right) {D} }
  }{
    \left(
      \left( 2 {m} + {\mu} \right) {D}
    \right) !
  }
  ,
\end{align}
${\alpha}\in\mathbb{R}_{+}$ is the cat code radius (the magnitude for coherent states equidistant around a circle in phase space) and ${D}\in\mathbb{N}_{>0}$ is the code distance, so that ideally the cat code can perfectly correct for up to ${D}-1$ lost photons.
${D}$ is half the number of coherent state components and also the separation between Fock state components with non-zero population in an arbitrary logical state.

Cat codes may be stabilised passively via reservoir engineering
\cite{mirrahimiDynamicallyProtectedCatqubits2014,albertHolonomicQuantumControl2016},
but for large ${D}$ the associated $2{D}$-photon interactions become increasingly difficult to produce.
We will be interested here in active measurement-based correction, specifically the tele-correction scheme
\cite{glancyTransmissionOpticalCoherentstate2004,liCatCodesOptimal2017,grimsmoQuantumComputingRotationSymmetric2020}.

\subsection{Phase Measurement}
\label{sub:phase_measurement}

\begin{figure*}[ht!]
  \includegraphics[width=0.99\textwidth]{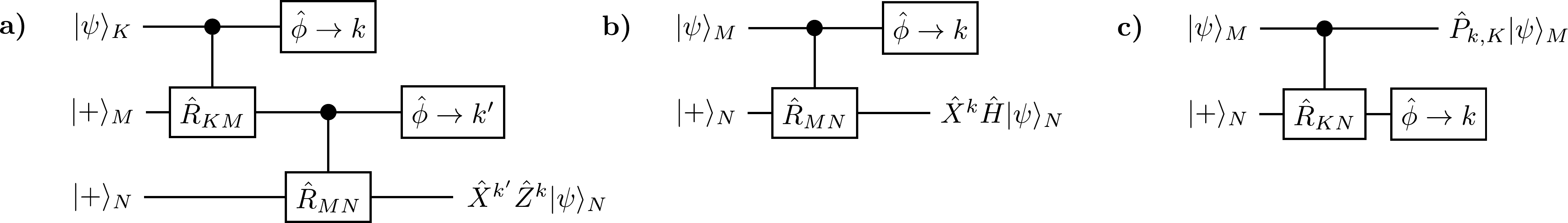}
  \caption{
      \label{fig:cat_tele_recovery_circuit}
      \textbf{a)}
      Tele-correction circuit for rotationally symmetric Bosonic codes
      \cite{grimsmoQuantumComputingRotationSymmetric2020},
      including cat codes, for arbitrary encoded input state ${\lvert\psi\rangle_{K}}$.
      Here the state subscripts ${K,M,N}$ are even and denote the numbers of coherent state components in the encoding.
      $\hat{R}_{MN}$ represents the controlled rotation gate:
      $e^{i\left(4\pi/MN\right)\hat{n}_{1}\hat{n}_{2}}$.
      Operators $\hat{\phi}$ denote general phase measurements, binned into indices ${k,k{'}}$ by the nearest coherent state components
      --- the pure output states shown here assume that components are perfectly distinguishable.
      \textbf{b)}
      Projective Hadamard circuit for rotation codes, which is repeated to form the full tele-correction circuit.
      \textbf{c)}
      Modular photon number measurement. Here ${\hat{P}_{k,K}}$ represents projection onto the subspace of photon numbers congruent to ${k}$ modulo ${K}$,
      and the phase angle in this case is binned into ${K}$ segments modulo ${4\pi/N}$.
  }
\end{figure*}

The tele-correction circuit, shown in Figure~\ref{fig:cat_tele_recovery_circuit}~(a),
can be used in the generation of rotationally symmetric Bosonic code states
\cite{grimsmoQuantumComputingRotationSymmetric2020},
including cat code states.
This circuit involves measurement of the phase of the data qubit ${\lvert\psi\rangle_{K}}$,
{
which provides information not only about the logical state but also about perturbations due to error.

Measurement inefficiencies have been highlighted as a dominating source of noise in teleportation-based error correction circuits
\cite{hillmannPerformanceTeleportationBasedErrorCorrection2022}.
It was recently shown by Oh et al.
\cite{ohOptimalGaussianMeasurements2019}
that a homodyne measurement is the optimal Gaussian measurement for the phase estimation problem, achieving a phase-independent sensitivity bound set by the quantum Fisher information. However, this measurement requires knowledge of an optimal quadrature angle, whereas multi-component cat code states can have an arbitrary degree of rotational symmetry.
For the balanced heterodyne measurement, the angular dependence is removed, but at the cost of halving the Fisher information, and so doubling corresponding mean-squared error.
While recent papers have considered adaptive homodyne measurements as an alternative
\cite{martinImplementationCanonicalPhase2020},
when the coherent state components of the code state are sufficiently separated
heterodyne detection nonetheless remains as a practical choice offering a favourable compromise between performance and implementability.
}

\begin{figure}[t]
  \begin{center}
    \includegraphics[width=0.39\textwidth]{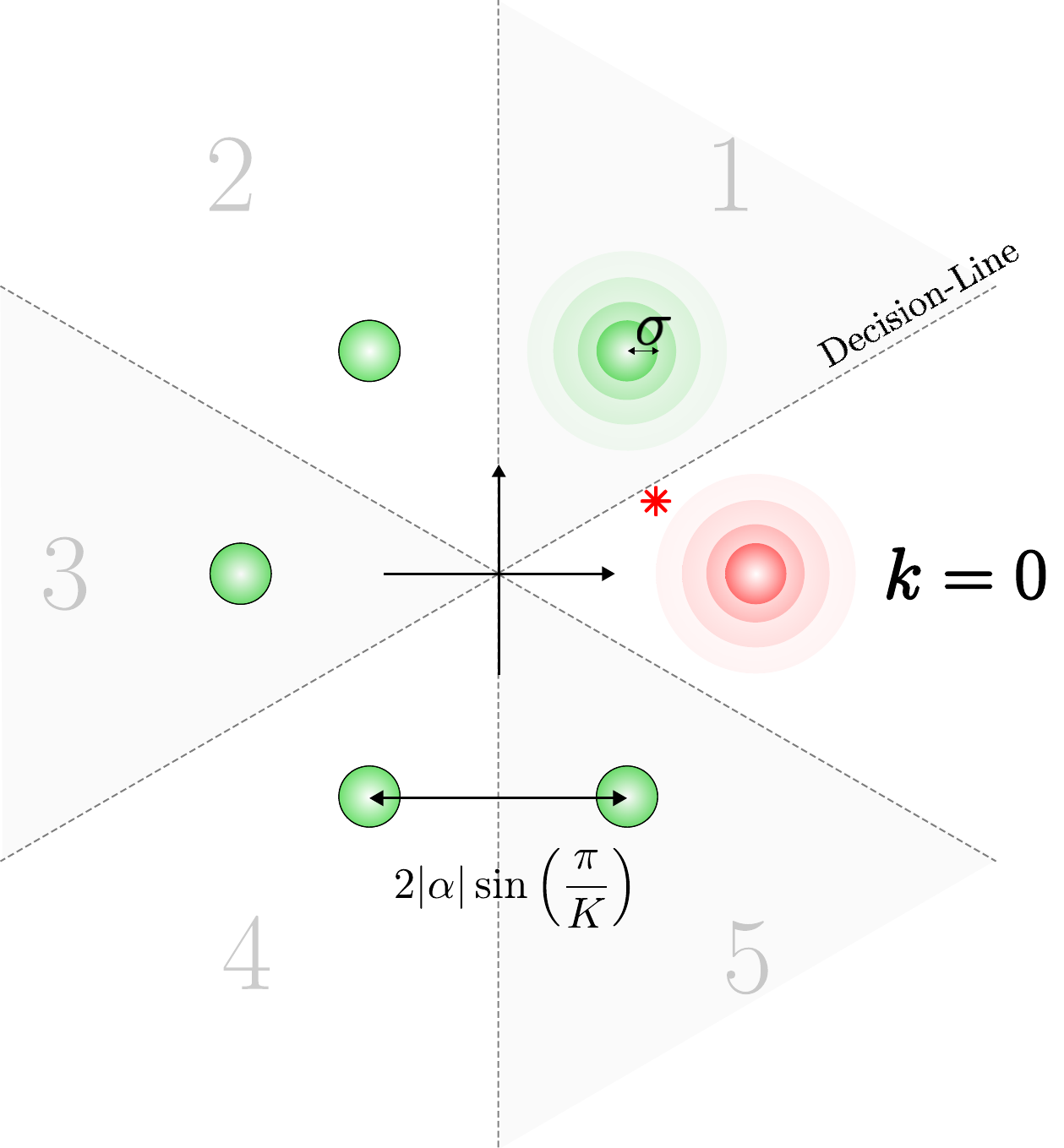}
  \end{center}
  \caption{
    \label{fig:cat_project_distance_sketch}
    Sketch of the binning procedure for heterodyne phase measurements in the tele-correction circuit of Figure~\ref{fig:cat_tele_recovery_circuit}~(a), with the number of components ${K=6}$ and ${k\in\{0:K-1\}}$.
    For example, if the location indicated by heterodyne detection is given by the red asterisk, the nearest coherent state component would be represented by the red concentric circles, corresponding to index ${k=0}$.
  }
\end{figure}

The ancillary states used in the tele-correction procedure of Figure~\ref{fig:cat_tele_recovery_circuit}~(a) need not be ${\lvert + \rangle}$ states.
Suppose that, for known ${\theta}$ and ${m}$, an ancillary state differs from the logical ${\lvert + \rangle}$ state by a rotation ${e^{i\theta \hat{n}}}$ about the origin and by an ${m}$-photon loss event ${\hat{a}^{m}}$.
The anomalous rotation commutes with the controlled rotation gates of the tele-correction circuit, and can be accounted for by a rotation of the decision lines for the heterodyne measurements (see Figure~\ref{fig:cat_project_distance_sketch}).
The anomalous photon loss event, affecting only the relative phases between coherent state components, does not affect the measurement outcomes directly. It does however result in a correlated rotation for other ancillae on commutation through a controlled rotation gate.
So long as ${m}$ is a known parameter, the angle of this correlated rotation is also known and can be accounted for as mentioned above.
These considerations amount to a qudit generalisation of Pauli-frame tracking, which has already been proposed for rotationally symmetric Bosonic codes \cite{grimsmoQuantumComputingRotationSymmetric2020} to allow for the random outcomes of the tele-correction and Hadamard circuits.

Assuming the code distance is large enough for heterodyne measurements to give accurate results, the final state of the tele-correction circuit in Figure~\ref{fig:cat_tele_recovery_circuit}~(a) is approximately
\begin{align}
  &
  \frac{c_{0}}{\sqrt{2}}
  \left(
    \vert + \rangle_{N}
    +
    e^{ - i \pi {k{'}} }
    \vert - \rangle_{N}
  \right)
  \nonumber\\
  &\qquad\qquad\qquad
  +
  \frac{
    c_{1}
    e^{ - i \pi {k} }
  }{\sqrt{2}}
  \left(
    \vert + \rangle_{N}
    -
    e^{ - i \pi {k{'}} }
    \vert - \rangle_{N}
  \right)
  ,
\end{align}
where the input state is ${\lvert\psi\rangle_{K} = c_{0} \lvert0\rangle_{K} + c_{1} \lvert1\rangle_{K}}$, while $k$, $k{'}$ index the coherent state components inferred from the measurement outcomes, as sketched in Figure~\ref{fig:cat_project_distance_sketch}.
An incorrect outcome in the first measurement would give an incorrect result for ${k}$, and likewise in the second measurement for ${k{'}}$.
If these incorrect outcomes differ from the genuine result by an odd amount, then the inferred and actual states are orthogonal.
False measurement outcomes must be suppressed by increasing ${\alpha}$ relative to ${K}$ and ${M}$ (the circumferential separation between coherent state components).
{
  An approximation to the error of such a measurement is described in Appendix~\ref{sec:measurement_error_through_total_variation_distance}.
}

\section{Many-Component Cat States}
\label{sec:many_component_cat_states}

Let us now look at generating high fidelity ancillary states using the Hadamard and modular photon number measurement circuits respectively depicted in Figure~\ref{fig:cat_tele_recovery_circuit}~(b)~and~(c).
Since $\vert + \rangle_{M}$ corresponds to $\vert 0 \rangle_{M/2}$
\cite{grimsmoQuantumComputingRotationSymmetric2020},
the simplest procedure
is to use two input coherent states with controlled rotation
$e^{i\left(4\pi/K\right)\hat{n}_{1}\hat{n}_{2}}$
in the modular photon number measurement circuit.
This gives
${\hat{a}^{K/2-k} \lvert + \rangle_{K}}$.
{
When ${K}$ is small however, generating a non-linear interaction of sufficient strength to enact this controlled rotation may not be feasible.
To address this issue, we introduce an ancillary many-component state generated by the Yurke--Stoler method.
Such a state may be prepared either by cross-Kerr interaction as described above or by self-Kerr non-linearity in a single field.
}
{
The generation of ${J}$-component states for this purpose (for ${J}$ large) allows us to simulate modular photon number measurements for small ${K}$ using interactions scaling as ${(KJ)^{-1}}$. This method of cat-like state generation is particularly suited to this purpose as the interaction strength required decreases as the number of components is increased, and this number is independent of ${K}$.
The Yurke--Stoler method of generating multi-component cat states has been considered before in a more general context
\cite{glancyMethodsProducingOptical2008},
but to our knowledge this is the first consideration of the impacts of the relative phases between their coherent state components and of loss on the subsequent measurements necessary for cat code tele-correction.
}

\subsection{Generalised Yurke--Stoler Method}
\label{sub:generalised_yurke_stoler_method}

In $1994$, Lee et al.
\cite{leeAmplificationMulticomponentSuperpositions1994}
generalised the method of Yurke and Stoler
\cite{yurkeGeneratingQuantumMechanical1986}
to generate multi-component circular cat states using Kerr non-linearities.
The Hamiltonian required for this method,
\begin{align}
  \hat{H}
  &=
  \hbar {\omega} \hat{{n}}
  +
  \hbar {\lambda} \hat{{n}}^{2}
  ,
\end{align}
where $\hat{{n}}$ is the photon number operator, involves a quadratic self-Kerr interaction in addition to the standard free energy of the mode.
Lee et al. noted that at times ${t}$ such that
\begin{align}
  {t}
  &=
  \frac{\pi}{{\lambda}{N}}
  ,
\end{align}
an ${N}$-component circular superposition of coherent states is obtained.
This superposition takes the form
\begin{align}
  \vert \Psi \rangle
  &=
  \frac{1}{\sqrt{\mathcal{N}}}
  \sum^{{N}}_{{m}=1}
  e^{i {\xi}_{{m}}}
  \vert - {\alpha} e^{i {\phi}_{{m}}} \rangle
  ,
\end{align}
where $\mathcal{N}$ is a normalisation constant, $e^{i {\phi}_{{m}}}$ are the ${N}$th roots of unity and the phases ${\xi_{m}}$ are described below.
We will refer to these as Yurke-Stoler (YS) states.
These states have been considered in prior work for the general creation of bosonic cat states
\cite{glancyMethodsProducingOptical2008}.
The Hamiltonian only contains terms that preserve the photon number distribution.
Consequently, the YS states do not have the modular photon number distribution characteristic of rotationally symmetric code states ---
the relative phases $e^{i {\xi}_{{m}}}$ are ``incorrect''.

The relative phases of the coherent state components using the generalised YS method of generation are solutions to the equation
\begin{align}
  \frac{1}{\sqrt{{N}}}
  \sum^{{N}}_{{m}=1}
  x_{{m}}
  e^{i 2\pi {k} \frac{{m}}{{N}}}
  &=
  e^{
    -i \pi \left( \frac{{k}^{2}}{{N}} + {k} \right)
  }
  ,
\end{align}
with
${x_{{m}} = e^{i {\xi}_{{m}}}}$.
We recognise that the relative phases then are the arguments of solutions to the discrete Fourier transform for the equation
\begin{align}
  e^{-i \pi \left( \frac{{k}^{2}}{{N}} + {k} \right) }
  ,
\end{align}
which we identify as a linear chirp-signal
\cite{smithDigitalSignalProcessing2003}.
The solution to the discrete Fourier transform gives us
\cite{berndtDeterminationGaussSums1981}
\begin{align}
  {\xi}_{{m}}
  &=
  \frac{\pi}{4}
  \left[
    {N}
    \left( \frac{2{m}}{{N}} - 1 \right)^{2}
    - 1
  \right]
  ,
  \label{eq:ys_relative_phases}
\end{align}
with ${{m} \in \{0,1,\ldots,{N}-1\}}$.
It can be readily checked that for the cases ${{N}\in\{2,3,4,5\}}$, these solutions correspond to those listed in
\cite{leeAmplificationMulticomponentSuperpositions1994}.

{

\subsection{The Role of YS-States in the Full Tele-correction Circuit}
\label{sub:the_role_of_the_yurke_stoler_states_in_the_full_telecorrection_circuit}

Our overall goal is to generate ancillary states to take the places of the ${\lvert + \rangle}$ states in the standard tele-correction circuit (Figure~\ref{fig:cat_tele_recovery_circuit}).
We propose to do this in two ways. First, for the third rail, we claim that the ${\lvert + \rangle_{N}}$ state itself can be generated from a YS-state and a coherent state using the modular photon number measurement described in Section~\ref{sub:modular_photon_number_measurement}.
It would in theory be possible to produce both ${\lvert + \rangle}$ states in this manner, but in
Section~\ref{sub:relative_phases_in_tele_correction} we observe that the state in the second rail can be replaced directly by an ${M/2}$ component YS-state, without using this YS-state to first generate ${\lvert + \rangle_{M}}$.
A consequence of using the YS-state directly in the tele-correction circuit is that the unwanted relative phases between coherent state components spread through the circuit, affecting measurement results and affecting the final output state. The self- and cross-Kerr corrections that are needed to account for this are the main focus of Section~\ref{sub:relative_phases_in_tele_correction}.

Putting together these approaches to the input states of the second and third rails, and denoting by ${\hat{n}^{2}_{M/2}}$ the self-Kerr evolution required to generate an ${M/2}$-component YS-state from an input coherent state, the total circuit is shown below in
Figure~\ref{fig:full_telecorrection_circuit_exploded_ys_state_generation}. The inputs to the second, third and fourth rails are all understood to be coherent states.

\begin{figure*}[ht]
  \centering
  \includegraphics{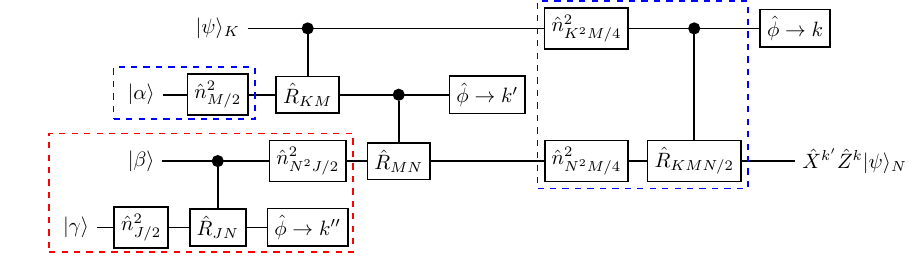}
  \caption{
    {
    The full, modified tele-correction circuit.
    Preparation of ${\lvert + \rangle}$ ancillary states has been replaced by YS-state preparation from coherent states and modular photon number measurement, and the resulting necessary quadratic corrections to relative phases have been added.
    ${\hat{n}^{2}_{K}}$ denotes the self-Kerr evolution required to generate an ${K}$-component YS-state.
    The magnitude of the nonlinearity required for each such phase correction scales inversely as the cube of the number of coherent state components typical for an input state.
    }
    \label{fig:full_telecorrection_circuit_exploded_ys_state_generation}
  }
\end{figure*}

}

{

\section{Propagation of Phases for State Initialisation}
\label{sec:propagation_of_phases_for_state_initialisation}

\subsection{The Transfer of Relative Phases in Controlled Rotation Gates}
\label{sub:the_transfer_of_relative_phases_in_controlled_rotation_gates}

Suppose we have two states, expressed respectively as superpositions of Fock states and a finite number of coherent states (of equal intensity and equispaced about the origin), of the form
\begin{align}
  \lvert \psi \rangle
  &=
  \sum^{\infty}_{n=0}
  c_{n}
  \lvert n \rangle
  ,\quad
  \lvert \phi \rangle
  =
  \sum^{N}_{j=1}
  d_{j}
  \lvert \alpha e^{i \frac{2\pi}{N} j} \rangle
  .
\end{align}
Applying a controlled rotation of angle ${\frac{2\pi}{KN}}$ between these two states results in
\begin{align}
  \sum^{\infty}_{n=0}
  \sum^{N}_{j=1}
  c_{n}
  d_{j}
  \lvert n , \alpha e^{i \frac{2\pi}{KN} (n + Kj)} \rangle
  .
\end{align}
Next, supposing ${\alpha \gg KN}$ and performing an approximate phase measurement on the second mode (collapsing the state to a single coherent state component), a measurement result consistent with a coherent state component at angle ${\frac{2\pi}{KN}M}$ (where ${M\in \{0,1,\ldots,KN-1\}}$) must satisfy
\begin{align}
  M
  &
  \cong
  n + Kj
  \quad
  (
    \text{modulo}
    \:
    KN
  )
  ,\nonumber\\
  M
  &\cong
  n
  \quad
  (
    \text{modulo}
    \:
    K
  )
  .
\end{align}
This measurement result leaves the residual state
\begin{align}
  \sum^{\infty}_{n\cong M (\text{mod}\: K)}
  c_{n}
  d_{j_{n}}
  \lvert n \rangle
  ,
  \quad
  j_{n}
  =
  \frac{M-n}{K} (\text{mod}\: N)
  ,
\end{align}
where we know ${\frac{M-n}{K}}$ must be an integer because
${M \cong n \: ( \text{mod} \: K )}$,
and the initial range ${j\in \{1,\ldots, N\}}$ ensures there is exactly one index ${j_{n}}$ satisfying this equation.
The index ${j_{n}}$ decrements sequentially (and periodically, with period ${N}$) in unit increments as ${n}$ increases sequentially in increments of ${K}$ in the remaining terms of the sum.
These predictable, sequential increments allow for post-measurement correction of undesired relative-phases in the coefficients ${d_{j_{n}}}$.

In this work, the coefficients ${d_{j}}$ have the form
${ d_{j} = e^{i \xi_{j} } ,}$
with ${\xi_{j}}$ defined in Equation~\eqref{eq:ys_relative_phases}.
These coefficients are periodic in ${j}$ (with period ${N}$), so that the now-superfluous ``modulo ${N}$'' condition in the above equation can be dropped, leaving simply
\begin{align}
  \sum^{\infty}_{n\cong M (\text{mod}\: K)}
  c_{n}
  e^{i \xi^{\prime}_{n} }
  \lvert n \rangle
  ,
\end{align}
where
\begin{align}
  {\xi}^{\prime}_{{n}}
  &=
  \frac{\pi}{4}
  \left[
    {N}
    \left( \frac{2{(M-n)}}{{K}{N}} - 1 \right)^{2}
    - 1
  \right]
  .
\end{align}
The ${M}$-dependence requires only a linear correction to the angle (up to a known global phase).
Extracting the strictly quadratic term, we have
${\frac{\pi{n^{2}}}{{K}^{2}{N}}}$,
and so the angle of the required self-Kerr correction scales inversely with ${{K}^{2}{N}}$.
}

\subsection{Modular Photon Number Measurement}
\label{sub:modular_photon_number_measurement}

{
In this section, we describe the application of the YS states to modular photon number measurement.
We will use the terms `probe' and `target' to refer to the measured and output rails, respectively.
}
In the modular photon number measurement of Figure~\ref{fig:cat_tele_recovery_circuit}~(c), we are able to account for relative phases between coherent state {components} in the probe subsystem. Using a coherent state ${\lvert\beta\rangle}$ as the target {input} subsystem and a YS state of amplitude ${\alpha}$ as the probe, we have
\begin{align}
  \frac{1}{\sqrt{\mathcal{N}}}
  \sum^{{N}}_{{m}=1}
  e^{i {\xi}_{{m}}}
  e^{-\frac{\lvert\beta\rvert}{2}}
  \sum^{\infty}_{n=0}
  \frac{\beta^{n}}{\sqrt{n!}}
  \lvert n\rangle
  \vert - {\alpha} e^{i {\phi}_{{m}}} e^{-i 4\pi \frac{n}{K N}} \rangle
\end{align}
after applying the controlled rotation gate.
We observe that the relative phases on the coherent state components of the YS state are transferred sequentially to the Fock states in the modular photon number subspace,
{
as described in Section~\ref{sub:the_transfer_of_relative_phases_in_controlled_rotation_gates}.
}
A second application of the self-Kerr interaction,
\begin{align}
  e^{i\frac{4\pi}{K^{2} N}\hat{n}^{2}}
  ,
\end{align}
removes the quadratic dependence in ${\xi_{m}}$.
The time required for this interaction is on the order of that for the generation of the YS states, and becomes shorter as the spacing between Fock states increases.

\begin{figure*}[ht!]
  \begin{center}
      \includegraphics[width=0.49\textwidth]{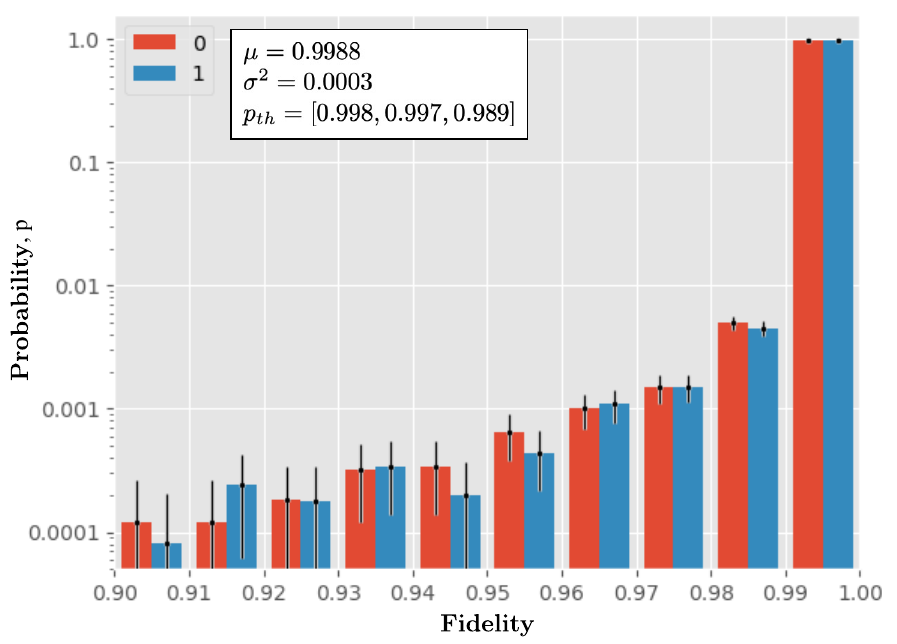}
      \includegraphics[width=0.49\textwidth]{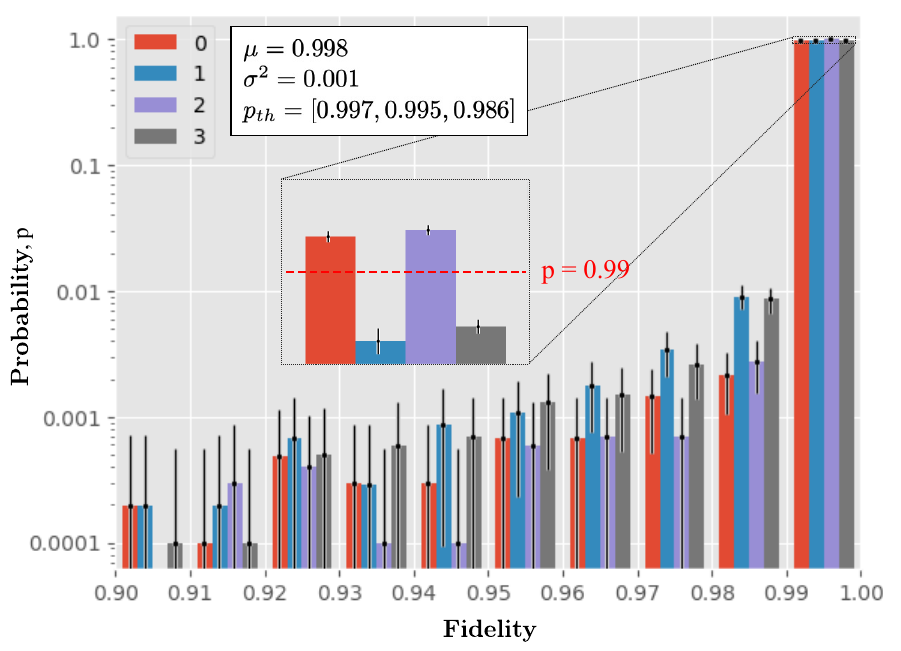}
      \caption{\label{fig:amplification_circuit_sample_results}
          Modular Photon Number Measurement shown in Figure~\ref{fig:cat_tele_recovery_circuit}~(c).
          Final state fidelities
          and their probabilities, with a probe cat state and a target coherent state of double amplitude.
          {Fidelities are binned into histograms to indicate success rates when post-selecting based on measurement outcome.}
          Histogram bars correspond to measurement outcome indices, as sketched in Figure~\ref{fig:cat_project_distance_sketch}.
          Error bars represent Wilson score ${95\%}$ confidence intervals.
          Threshold probabilities
          {${p_{th}}$}
          are given for fidelities above ${0.9}$, ${0.95}$ and ${0.99}$ respectively,
          {alongside the mean fidelity ${\mu}$ and its variance ${\sigma^{2}}$.}
          \textbf{(Left)}
          {Measuring the photon number ${\mathrm{mod}\:2}$, using}
          a probe cat state with ${2}$ coherent state components and ${\alpha=4}$, and heterodyne local oscillators with amplitude ${\beta=6}$.
          ${10^{5}}$ samples taken.
          \textbf{(Right)}
          {Measuring the photon number ${\mathrm{mod}\:4}$, using}
          a probe cat state with ${4}$ coherent state components and ${\alpha=12}$, and heterodyne local oscillators of amplitude ${\beta=8}$.
          ${4\times 10^{4}}$ samples taken.
          \textbf{(Inset)}
          The ${4}$-component results display a statistically significant difference in high-fidelity probabilities for even and odd indices, located either side of ${p=0.99}$, {a possible artifact of the small choice of ${\beta}$ and discretised grid of measurement outcomes.}
      }
  \end{center}
\end{figure*}

{
To demonstrate the stability of the modular photon number measurement against measurement error as we increase the number of coherent state components, in Figure~\ref{fig:amplification_circuit_sample_results} we sample measurement outcomes and the final state fidelities with a coherent state target {input} and cat state probe.
{Measurement outcomes are sampled randomly for circuits implementing photon number measurements ${\mathrm{mod}\:2}$ and ${\mathrm{mod}\:4}$.}
Final states with fidelities above ${0.99}$ are observed with probability ${p > 0.985}$.
The fidelity
\begin{align}
  F
  &=
  \lvert \langle \psi_{\mathrm{actual}} \vert \psi_{\mathrm{target}} \rangle \rvert^{2}
\end{align}
is taken with respect to the most likely final cat state ${\hat{a}^{K/2-k} \lvert + \rangle_{K}}$, not restricted to ${\lvert + \rangle}$.
}

\subsection{Relative Phases in Tele-correction}
\label{sub:relative_phases_in_tele_correction}

While the relative phases of the YS states can be corrected in modular photon number measurements, it may be expected that they would remain detrimental in tele-correction as these states do not have the photon number distribution of true cat code states.

\begin{figure}[t]
  \begin{center}
      \includegraphics[width=0.49\textwidth]{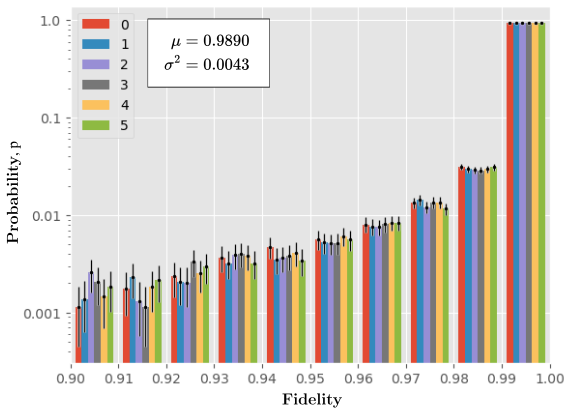}
      \caption{\label{fig:ys_in_central_rail}
        {
        Final state fidelities and their probabilities for the
        tele-correction as shown in Figure~\ref{fig:cat_tele_recovery_circuit}~(a), with a Yurke-Stoler input state to the second rail.
        The first and third rails are initialised with ideal ${\lvert +\rangle_{6}}$ states, with ${\lvert\alpha\rvert=4}$.
        The second rail uses the Yurke--Stoler state initialisation to simulate preparation of a ${\lvert +\rangle_{12}}$ state with ${\lvert\alpha\rvert=8}$.
        ${80000}$ samples have been taken, and error bars represent Wilson score ${95\%}$ confidence intervals.
        Fidelities are binned into histograms to indicate success rates when post-selecting based on measurement outcome.
        Clustered histogram bars correspond to measurement outcome indices, as sketched in Figure~\ref{fig:cat_project_distance_sketch}.
        The probabilities of achieving fidelities above
        ${\{0.9, 0.95, 0.99\}}$ are
        ${\{ 0.977, 0.963, 0.908 \}}$ respectively.
        The mean output fidelity ${\mu=0.989}$ and variance ${\sigma^{2}=0.0043}$ are also shown.
        }
      }
  \end{center}
\end{figure}

If the YS state is used as the input to the second rail of the tele-correction circuit, we may interpret the combination of both controlled rotations and the central measurement together as a measurement of the modular sum of photon numbers in the first and third rails. The resultant state in this case, prior to measurement, would be
\begin{widetext}
  \begin{align}
    &
    \frac{1}{\sqrt{\mathcal{N}}}
    \sum_{{n},{n{'}}}
    \left[
      \frac{
        {\beta}^{ {a}_{{n},{\mu}} {K}/2 }
      }{
        \sqrt{
          \left(
            {a}_{{n},{\mu}} {K}/2
          \right) !
        }
      }
      \frac{
        {\delta}^{ {b}_{{n{'}},{\nu}} {K}{'}/2 }
      }{
        \sqrt{
          \left(
            {b}_{{n{'}},{\nu}} {K}{'}/2
          \right) !
        }
      }
    \right]
    \vert {a}_{{n},{\mu}} {K}/2 \rangle
    \otimes
    \left[
      \sum^{{N}}_{{m}=1}
      e^{i {\xi}_{{m}}}
      \vert - {\alpha} e^{i {\phi}_{{m}}} e^{-i \frac{2\pi}{N} \left( {a}_{{n},{\mu}} + {b}_{{n{'}},{\nu}} \right)} \rangle
    \right]
    \otimes
    \vert {b}_{{n{'}},{\nu}} {K}{'}/2 \rangle
    ,
  \end{align}
\end{widetext}
for
${{a}_{{n},{\mu}} = 2 {n} + {\mu}}$, ${{b}_{{n{'}},{\nu}} = 2 {n{'}} + {\nu}}$,
some normalisation constant ${\mathcal{N}}$ and ${\mu,\nu\in\{0,1\}}$.
The relative phases are again inherited sequentially by the Fock states in the modular subspace,
{
as described in Section~\ref{sub:the_transfer_of_relative_phases_in_controlled_rotation_gates},
}
but in this case a cross-Kerr interaction
\begin{align}
  e^{i\frac{4\pi}{K^{2} N}\hat{n}^{2}_{1}}
  e^{i\frac{4\pi}{K{'}^{2} N}\hat{n}^{2}_{2}}
  e^{i\frac{8\pi}{K K{'} N} \hat{n}_{1} \hat{n}_{2}}
  ,
\end{align}
is required to remove the quadratic dependence.
The stable performance of the tele-correction scheme in this situation under measurement error is verified in the results of Figure~\ref{fig:ys_in_central_rail}.

{

\section{Noise and Logical Error}
\label{sec:noise_and_logical_error}

We next consider experimental feasibility from the perspective of gate-level noise sources.

Figure~\ref{fig:full_telecorrection_circuit_exploded_ys_state_generation} introduces three new subcircuits into the standard tele-correction procedure.
In addition to the common loss- and dephasing-noise models in the original structure, we must also consider how these subcircuits themselves might constrain performance.

We will consider four sources of noise in the control of the new subcircuits:
dephasing noise, non-linear over- or under-rotations, loss, and displacement noise at the coherent state inputs.
The first two of these sources, dephasing and non-linear rotations, are assumed to result primarily from effective timing error when implementing the non-linear self- and cross-Kerr gates.
The third, loss, is particularly relevant to the comparative discussion of non-linear materials in Appendix~\ref{sec:utility_thresholds_for_the_third_order_nonlinear_susceptibility} (indeed, the ratio between the loss-rate and non-linear index of refraction has been considered a key quantity in prior studies \cite{glancyMethodsProducingOptical2008}).
The final source, displacement noise, is a result specifically of our chosen method for producing superposed coherent states, which relies on the production of a high-quality coherent state input.

As will be shown below, discretised error models for dephasing, non-linear rotation, and loss noise sources can all be framed in such a way as to commute with the rotation gates in Figure~\ref{fig:full_telecorrection_circuit_exploded_ys_state_generation}.
The key performance metric that we will use to quantify the effects of these sources of noise is the probability that one of the heterodyne measurements in the first two rails gives an incorrect result (after the binning procedure depicted in Figure~\ref{fig:cat_project_distance_sketch}).
This is equivalent to a `packet error ratio', where the packets in this case correspond to the binary error information extracted by the tele-correction circuit.

\subsection{Dephasing Noise}
\label{sub:dephasing_noise_across_the_extended_circuit}

We begin with dephasing noise.
For the self-Kerr interaction,
\begin{align}
  e^{
    i
    \lambda
    \hat{n}^{2}
    \tau
    +
    i
    \omega
    \hat{n}
    \tau
  }
  ,
\end{align}
dephasing can arise due to imprecise control over the relative magnitudes of ${\lambda}$ and ${\omega}$ or in the interaction time ${\tau}$.
In the cross-Kerr interaction,
\begin{align}
  e^{
    i
    \lambda
    \hat{n}_{1}
    \hat{n}_{2}
    \tau
  }
  ,
\end{align}
on measuring the phase of one subsystem via heterodyne measurement, dephasing can occur due to imprecise control of the interaction time ${\tau}$, or due to prior erroneous changes in the photon number distribution of the measured subsystem (such as occurs due to loss).

One key observation that simplifies the analysis of this form of error is that it commutes with all gates in Figure~\ref{fig:full_telecorrection_circuit_exploded_ys_state_generation}, so that we can assess the error rate by looking directly at the impact of the phase rotation on the single measurement error rate.
To quantify the phase rotation error, we write
\begin{align}
  \hat{U}_{\text{err}}
  &=
  e^{
    i
    \frac{2\pi}{N}
    \theta
    \hat{n}
  }
  ,
\end{align}
taking the magnitude of the quantity ${\theta}$ to represent a relative error magnitude.
This quantity ${\theta}$ is expressed as an additional factor multiplying ${\frac{2\pi}{N}}$, where ${N}$ represents the number of coherent state components in the superposition, because all self- and cross-Kerr rotations include an exponent inversely proportional to ${N}$
(i.e. ${\theta}$ is expressed here as a relative error proportional to the total interaction time).

The sampled numerical results for the binned heterodyne measurement error across a range of values for ${N}$, ${\theta}$ and coherent state magnitude ${\lvert \alpha \rvert}$ are shown in Figure~\ref{fig:fractional_rotation_measurement_error}.
In this figure, ${\theta}$ is expressed as a (negative) power of ${2}$, so that changes can be conveniently expressed as integer increments on a log-scale as ${\log_{2}(\theta)}$.
As expected for this error model, the measurement error rate ${P_{\text{err}}}$ is suppressed exponentially as ${\lvert\alpha \rvert}$ increases, for all values of ${\theta}$.
We observe no minimum error rate, and at these scales performance appears unaffected by ${\theta}$ (within the expected sample error) for even the largest values considered
(${2^{-8}\sim 4\times 10^{-3}}$).
To explain this behaviour, we invoke an approximation to the error rate through the total variation distance described in Appendix~\ref{sec:measurement_error_through_total_variation_distance}.
An approximate expression for the error is
\begin{align}
  \frac{1}{2}
  e^{
    -\frac{1}{2}
    \left(
      2
      \lvert\alpha\rvert
      \sin\left(
        \frac{\pi}{\tilde{N}}
      \right)
    \right)
  }
  ,\quad
  \tilde{N}
  \equiv
  N \left( 1 - \theta \right)
  ,
\end{align}
where ${\tilde{N}}$ represents an effective number of components for the purpose of calculating the error rate, determined by the relative reduction ${1\rightarrow 1 - \theta}$ in the angular spacing between components.
At the upper limit for the range of ${\theta}$ considered, the \emph{relative} increase in the expected error as a result of the transformation
${N\rightarrow \tilde{N}}$ ranges between ${10^{-3}}$ and ${0.15}$ as ${\lvert\alpha\rvert}$ increases, comparable at all points with statistical fluctuations.
\begin{figure}[ht]
  \centering
  \includegraphics[width=0.49\textwidth]{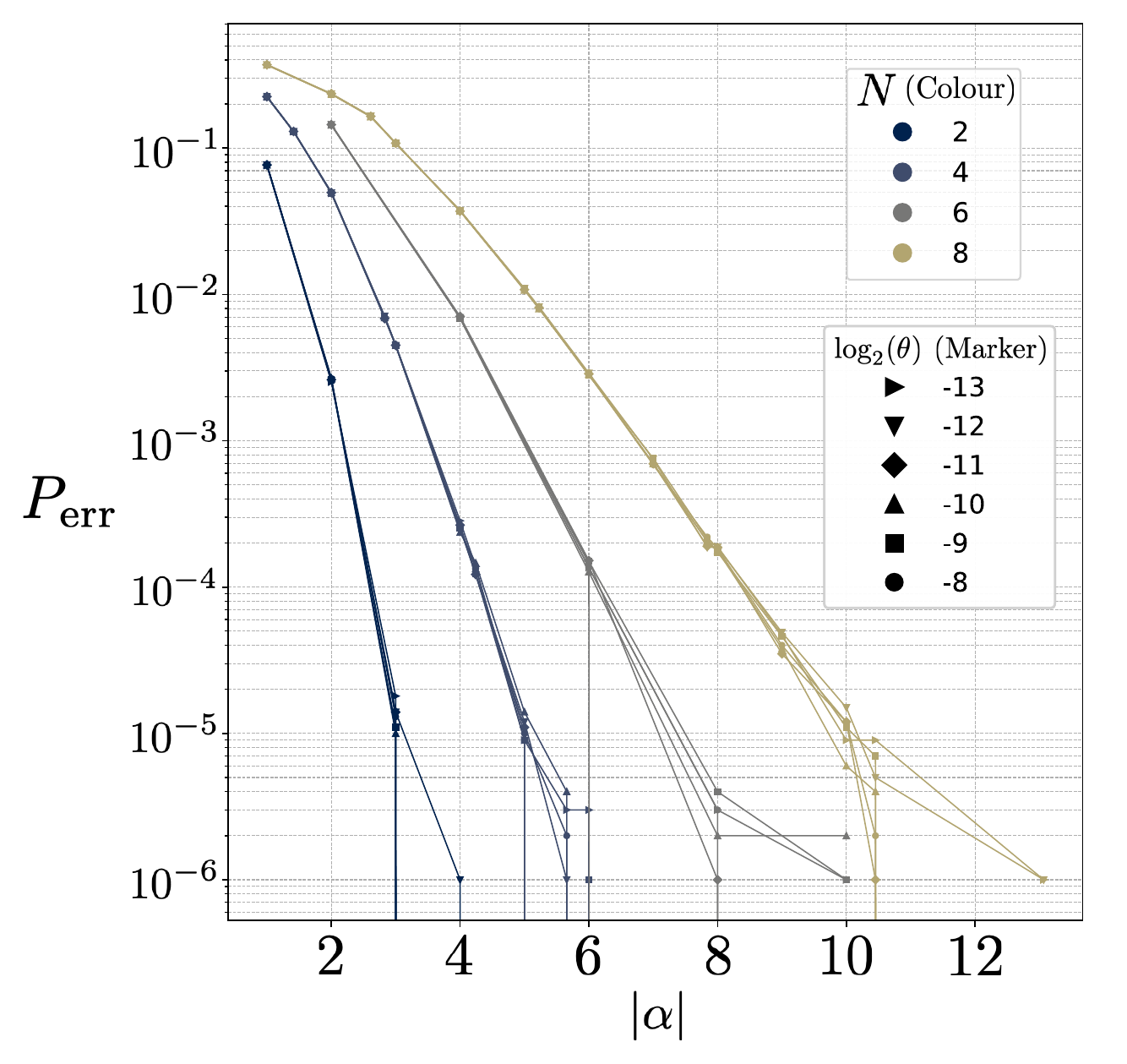}
  \caption{
    \label{fig:fractional_rotation_measurement_error}
    {
    Binned heterodyne detection measurement error ${P_{\text{err}}}$ as a function of the number of superposed coherent state components ${N}$, the coherent state magnitude ${\lvert \alpha \rvert}$, and the magnitude ${\theta}$ of an erroneous phase rotation
    ${e^{i \frac{2\pi}{N} \theta \hat{n} }}$.
    Exponential decay in the error rate with ${\lvert \alpha \rvert}$ is observed, with no lower bound, for all parameters.
    ${\theta}$ is drawn from the set ${\{2^{-13}, 2^{-12}, 2^{-11}, 2^{-10}, 2^{-9}, 2^{-8}\}}$, expressed in the legend on a log-scale for clarity.
    For ${\theta}$ in this range, the increase in the measurement error rate due to the erroneous phase rotation remains comparable to sample error.
    ${10^{6}}$ samples taken for each point. Sample error follows a binomial distribution.
    }
  }
\end{figure}

\subsection{Non-linear Rotation Noise}
\label{sub:non_linear_rotation_noise_across_the_extended_circuit}

We next consider non-linear rotation error.
This can arise in self-Kerr interactions
\begin{align}
  e^{
    i
    \lambda
    \hat{n}^{2}
    \tau
  }
  ,
\end{align}
as imprecision in either the magnitude of ${\lambda}$ or the interaction time ${\tau}$.

Just as for the simpler dephasing channel, this form of error commutes with all gates in Figure~\ref{fig:full_telecorrection_circuit_exploded_ys_state_generation}.
We therefore again consider its effect directly on the error rate of the binned heterodyne measurements.

To quantify the non-linear rotation error, we write
\begin{align}
  \hat{U}_{\text{err}}
  &=
  e^{
    i
    \frac{2\pi}{N}
    \theta
    \hat{n}^{2}
  }
  ,
\end{align}
taking the magnitude of the quantity ${\theta}$ to represent a relative error magnitude, where ${N}$ represents the number of coherent state components.
The motivation behind this choice for ${\theta}$ follows that for the dephasing channel: all self-Kerr rotations include an exponent inversely proportional to ${N}$, and so ${\theta}$ is a relative error proportional to the total interaction time.

The sampled numerical results for the binned heterodyne measurement error across a range of values for ${N}$, ${\theta}$ and coherent state magnitude ${\lvert \alpha \rvert}$ are shown in
Figure~\ref{fig:self_kerr_rotation_measurement_error}.
As in Subsection~\ref{sub:dephasing_noise_across_the_extended_circuit}, ${\theta}$ is expressed as a (negative) power of ${2}$.
For small ${\lvert \alpha \rvert}$, exponential decay is again observed as the components initially become distinguishable from one another. Unlike in the case of simple dephasing however, the angle of rotation for this non-linear gate depends explicitly on the photon number, the average of which increases as ${\lvert\alpha\rvert^{2}}$. The result is that a ${\theta}$-dependent minimum error rate is observed, followed by an exponential increase in the error rate until this rate reaches ${\mathrm{O}(1)}$.
We observe that for a threshold fidelity in the range ${0.999}$ to ${0.99}$, the four-component state can tolerate ${\theta}$ up to ${2^{-9}}$ or ${2^{-8}}$ (${\sim4\times 10^{-3}}$), whereas the eight-component case can tolerate ${\theta}$ only up to ${2^{-11}}$ or ${2^{-10}}$ (${\sim10^{-3}}$).
\begin{figure}[ht]
  \centering
  \includegraphics[width=0.49\textwidth]{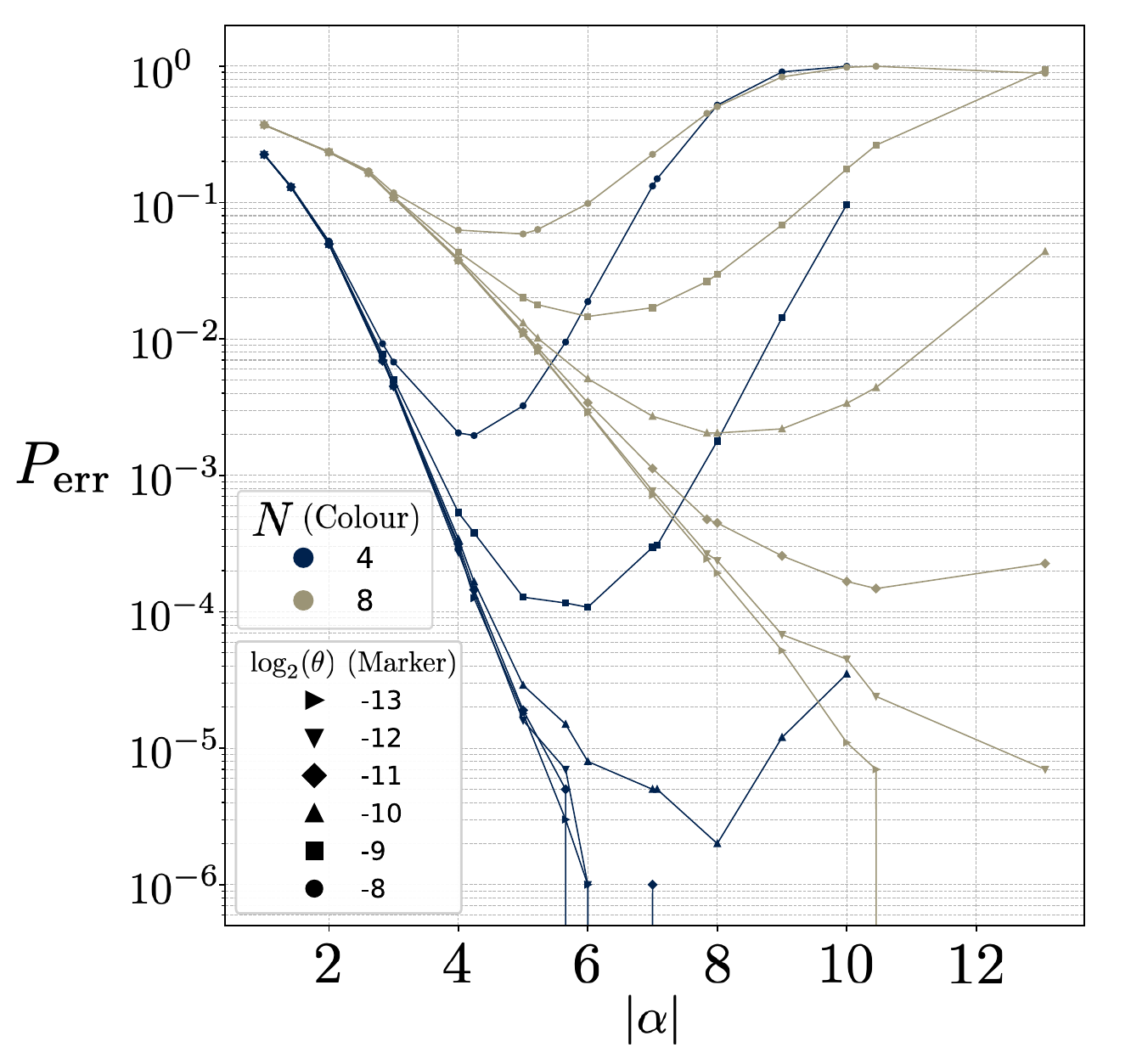}
  \caption{
    \label{fig:self_kerr_rotation_measurement_error}
    {
    Binned heterodyne detection measurement error ${P_{\text{err}}}$ as a function of the number of superposed coherent state components ${N}$, the coherent state magnitude ${\lvert \alpha \rvert}$, and the magnitude ${\theta}$ of an erroneous nonlinear phase rotation
    ${e^{ i \frac{2\pi}{N} \theta \hat{n}^{2} }}$.
    Exponential decay in the error rate with ${\lvert \alpha \rvert}$ is observed, reaching a minimum value and then followed by an exponential increase.
    ${\theta}$ is drawn from the set ${\{2^{-13}, 2^{-12}, 2^{-11}, 2^{-10}, 2^{-9}, 2^{-8}\}}$, expressed in the legend on a log-scale for clarity.
    ${10^{6}}$ samples taken for each point. Sample error follows a binomial distribution.
    }
  }
\end{figure}

\subsection{Displacement Noise at the Input}
\label{sub:displacement_noise_at_the_input}

The Gaussian displacement channel for our input states describes random phase-space shifts applied to a single-mode coherent state, where the amplitude $\alpha$ of the state follows a distribution with variance $\sigma ^2= (\sigma^\prime |\alpha|)^2$ that scales with $|\alpha|^2$. Physically, random displacement can arise from thermalisation with the environment, imperfect laser sources (intensity and phase noise) as well as technical jitter in the optical setup.

The impact of displacement noise on our circuits is twofold. First, when propagating through non-linear components, the noise induces number-dependent effects. In self-Kerr interactions (during YS state generation), the channel becomes photon-number dependent, while in cross-Kerr interactions (during telecorrection), displacement noise rotates the phase conditioned on excitations in other rails. In heterodyne detection, it manifests as a broadening of the outcome distribution. Together, these effects lead to a small reduction in fidelity, depending on the chosen scaling $\sigma ^\prime$.

We estimate a reasonable magnitude for ${\sigma^\prime}$ in Appendix~\ref{sec:random_displacement_width_from_relative_intensity_noise} from assumptions about typical levels of the relative intensity noise of laser sources.
Choosing $|\alpha| = 6$, we estimate $\bar{n} \sim 1.825 \times 10 ^{-11}$, and conclude that a reasonable assumption for the scaling of amplitude due to displacement is on the order of
\begin{equation}
\begin{aligned}
    \sigma^\prime  &= \sigma/|\alpha | = \sqrt{\bar{n}/2}/|\alpha|   \\
    &\sim 5 \times 10^{-7}.
\end{aligned}
 \end{equation}

Mathematically, Gaussian-distributed random displacement noise can be expressed as the concatenation of a pure-loss channel $\hat{E}_l$ with an amplification channel $\hat{A}_l$ with some loss rate $\gamma$
\cite{albertPerformanceStructureSinglemode2018},
\begin{equation}
  \begin{aligned}
      \hat{E}_l &= (\frac{\gamma}{1-\gamma})^{l/2} \frac{1}{\sqrt{l!}} \hat{a}^l (1-\gamma) ^{\hat{n}/2}  \\
      \hat{A}_l &= \sqrt{1-\gamma } \hat{E}_l^\dagger, \\
  \end{aligned}
  \label{eq:loss_and_amplification_krauss_ops}
\end{equation}
where $l \geq 0$.
We simulate the effect of displacement noise in the input to the second rail in the tele-correction circuit by taking ideal logical ${\lvert + \rangle}$ states on the first and third rails, and implementing this concatenated channel on the coherent state input to the second rail, before then simulating the rest of the ideal tele-correction circuit and assessing the fidelity of the output state from the third rail (such fidelities are aggregated over ${10^{4}}$ samples, taking their mean).
We conservatively take ${\sigma ^\prime}$ to be strictly larger than our estimate above, across the range $\sigma ^\prime \in \{ 10^{-6}, 10^{-5}, 10^{-4}, 10^{-3}, 10^{-2} \}$.

The results, shown in Figure~\ref{fig:mean_fidelities_under_displacement_noise}, illustrate the effect of displacement noise on state fidelity as a function of the noise scaling parameter ${\sigma ^\prime }$. The mean fidelity remains above 0.992 for noise scales up to $\sigma^\prime \sim 10^{-3}$  with some statistical fluctuations, indicating that our protocol is robust against small levels of displacement noise. However, as the noise scaling increases to $\sigma^\prime \sim 10^{-2}$, the mean fidelity experiences a sharp decline.
While the protocol is in principle vulnerable to displacement noise in the input coherent states, it is expected to maintain high fidelity under reasonable noise assumptions.

For small displacements, we can take the matrix elements of the displacement operators in the number basis
\cite{cahillOrderedExpansionsBoson1969}
\begin{align}
  \langle
    m | \hat{D}(\alpha) | n
  \rangle
  &=
  e^{-|\alpha|^{2}/2}
  \sqrt{\frac{n!}{m!}}
  \alpha^{m-n}
  L_{n}^{(m-n)}(|\alpha|^{2})
  ,\nonumber\\
  m &\geq n
  ,
\end{align}
and expand the Laguerre polynomials to lowest-order in ${|\alpha|^{2}}$.
The result is proportional to
\begin{align}
  \sqrt{\frac{n!}{m!}}
  \alpha^{m-n}
  (m-n)^{n}
  ,
\end{align}
a rapidly decaying function of the photon number separation, which may be protected by increasing the code distance in the number basis.
A particular displacement applied to a coherent state may be expressed (up to normalisation) as
\begin{align}
  s^{\hat{a}^{\dagger}\hat{a}}
  \lvert\gamma \rangle
  &=
  \lvert s \cdot \gamma \rangle
  ,
\end{align}
for some ${s\in \mathbb{C}}$ that depends both on the initial coherent state and the displacement amplitude.
This operator commutes with all non-linear and rotation gates in the tele-correction circuit.
While the distribution over ${s}$ may not follow the same Gaussian distribution as the arguments ${\alpha}$ to the displacement operator, this commutation ensures that the tele-correction circuit remains fault-tolerant for bounded ${s}$.

\begin{figure}[ht]
  \centering
  \includegraphics[width=0.45\textwidth]{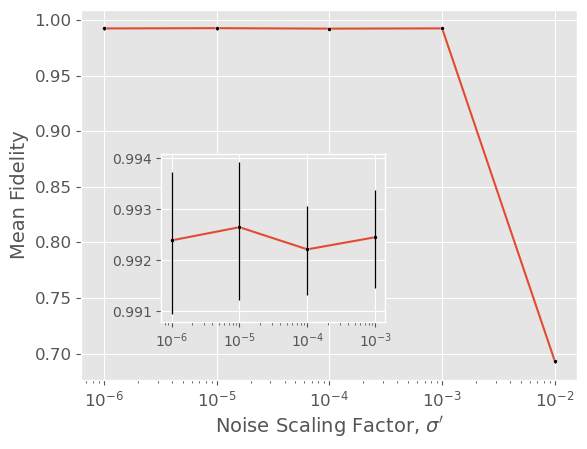}
  \caption{
    \label{fig:mean_fidelities_under_displacement_noise}
    {
    Mean fidelities for an otherwise ideal tele-correction circuit, when the input coherent state to the central rail is subject to Gaussian-distributed random displacement noise with variance ${\sigma ^2= (\sigma^\prime |\alpha|)^2}$ prior to the self-Kerr interaction.
    The relative standard deviations ${\sigma ^\prime}$ are drawn from the set
    ${\sigma ^\prime \in \{ 10^{-6}, 10^{-5}, 10^{-4}, 10^{-3}, 10^{-2} \}}$.
    For each ${\sigma^{\prime}}$ in this set, we have taken (listed in order)
    ${\{6400, 6400, 16040, 12800, 12800\}}$ samples.
    Ideal ${\lvert + \rangle_{4}}$ states are input to the first and third rails, with ${\lvert\alpha\rvert = 3}$ for these states.
    For the central rail, the self-Kerr interaction simulates preparation of a ${\lvert + \rangle_{8}}$ state with ${\lvert\alpha\rvert = 6}$.
    Mean output state fidelities are
    ${\{0.9924, 0.9926, 0.9922, 0.9925, 0.6930\}}$.
    The performance of the protocol remains stable up to ${\sigma ^\prime = 10^{-3}}$, but declines sharply as it approaches ${10^{-2}}$.
    }
  }
\end{figure}

\subsection{Loss during Yurke--Stoler State Initialisation}
\label{sub:loss_during_yurke_stoler_state_initialisation}

We next consider the effects of loss during the initialisation of the Yurke--Stoler states, beginning by considering loss events before and after the non-linear channel (as typically assumed in a discretised error treatment), and then discussing the detrimental impact of loss events occuring during the channel itself.

Firstly, suppose a zero-temperature loss channel (described by the Kraus operators in Equation~\eqref{eq:loss_and_amplification_krauss_ops}) is applied after coherent state initialisation but before the self-Kerr non-linearity.
Because the initial state is a coherent state (an eigenstate of ${\hat{a}}$), this channel is equivalent to applying the operator
\begin{align}
  \left(
    1 - \gamma
  \right)^{ \frac{1}{2} \hat{a}^{\dagger}\hat{a} }
  ,
\end{align}
which has the simple effect of reducing the coherent state amplitude.
The exponent ${\hat{a}^{\dagger}\hat{a}}$ commutes with all gates in the circuit, and so we may treat directly the impact on binned heterodyne measurement of this amplitude reduction.
We established in Figure~\ref{fig:fractional_rotation_measurement_error} how the error scales with ${\lvert\alpha\rvert}$ (in the absence of other error sources), so we can read directly from that figure that the measurement error will increase exponentially as ${(1-\gamma)^{\frac{1}{2}}}$ decreases, with a gradient proportional to the number of coherent state components. Such a reduction could be compensated by increasing the input coherent state amplitude.

Secondly, suppose a pure loss channel is applied after the self-Kerr interaction. After this point, the only changes in the extended version of the circuit in Figure~\ref{fig:full_telecorrection_circuit_exploded_ys_state_generation} as compared with the standard tele-correction circuit are the corrective operations at the end of the circuit.
These corrections do not affect the central rail however (they commute with this loss channel), so the error in the circuit should in this case be exactly the same as for the original tele-correction circuit.
A ${k}$-photon loss event occuring on the newly introduced fourth rail with coherent state ${\lvert\gamma\rangle}$ will manifest as an erroneous phase rotation on the third rail, of angle ${\frac{4\pi}{JN}\cdot k}$, and is correspondingly suppressed with increasing ${J}$.
The measurement error rates associated with such a rotation were considered in Figure~\ref{fig:fractional_rotation_measurement_error}.

Thirdly, more interesting behaviour emerges when photon loss events occur \emph{during} the self-Kerr interaction.
Suppose initially that the strength of the self-Kerr interaction is time independent.
In this scenario, and scaling the time variable by the interaction strength ${\lambda}$ so that the total scaled time of the interaction is
${\tau_{1} + \tau_{2} =1}$,
the effect of a single photon annihilation operator occuring at a time ${\tau_{1}}$ from the beginning of the interaction is
\begin{align}
  e^{
    i
    \frac{2\pi}{M}
    \tau_{2}
    \hat{n}^{2}
  }
  \hat{a}
  e^{
    i
    \frac{2\pi}{M}
    \tau_{1}
    \hat{n}^{2}
  }
  \lvert \alpha \rangle
  &=
  \alpha
  e^{
    i
    \frac{2\pi}{M}
    \tau_{2}
    \hat{n}^{2}
  }
  e^{
    i
    \frac{2\pi}{M}
    \tau_{1}
    (\hat{n} + 1)^{2}
  }
  \lvert \alpha \rangle
  ,\nonumber\\
  &=
  \alpha
  e^{
    i
    \frac{2\pi}{M}
    \tau_{1}
  }
  e^{
    i
    \frac{4\pi}{M}
    \tau_{1}
    \hat{n}
  }
  e^{
    i
    \frac{2\pi}{M}
    \hat{n}^{2}
  }
  \lvert \alpha \rangle
  .
\end{align}
Here ${\lvert \alpha e^{ i \frac{2\pi}{M} \tau_{1} } \rvert^{2}}$ weights the probability of the single-photon loss event, but the term
${ e^{ i \frac{4\pi}{M} \tau_{1} \hat{n} } }$
introduces an erroneous phase rotation that, depending on ${\tau_{1}}$,
rotates the state by an angle between ${0}$ and ${\frac{4\pi}{M}}$ in phase space.
Since there are ${M/2}$ components generated for this Yurke--Stoler state, if the distribution over ${\tau_{1}}$ is uniform, then the population becomes uniformly distributed about the unit circle.
Fault-tolerance breaks down in such a case, as for even a single-photon loss event angular changes due to interactions with the other rails can no longer be detected. This is independent of the amplitude ${\alpha}$ or the number of components ${M}$.

The probability of photon loss is proportional to the intensity of the pulse, while the effective self-Kerr interaction strength is proportional to its square.
Following the treatment developed in
\cite{drummondQuantumTheoryNonlinear2014,brechtPhotonTemporalModes2015},
we will introduce an effective non-linear interaction strength ${\chi_{\text{eff}}}$ proportional to the squared intensity,
\begin{align}
  \chi_{\text{eff}}
  &=
  \chi_{0}
  \int^{\infty}_{-\infty} d\tau
  I^{2}(\tau)
  .
\end{align}
For simplicity, we assume that the medium's nonlinear response is much faster than the pulse envelope (allowing for the definition of an instantaneous effective coupling ${\chi_{\text{eff}}(\tau) = \chi_{0} I^{2}(\tau) }$), and that effects such as dispersion and orthogonal-mode coupling are negligible.
Instead of the rectangular pulse implied by a time-independent self-Kerr interaction strength, consider now a pulse with a time-dependent intensity described by a scalar multiple of some absolutely continuous probability density function ${g(\tau)}$.
The instantaneous pulse intensity at a time ${\tau_{1}}$ is then ${I(\tau_{1}) = I_{0} g(\tau_{1})}$.
The effective Hamiltonian for the self-Kerr interaction is
\begin{align}
  \hat H(\tau) &=
  \hbar
  \frac{\chi_{\text{eff}}(\tau)}{2}
  \hat{n}^{2}
  ,
\end{align}
and the evolution operator over the full gate is
\begin{align}
  e^{
    -i
    \frac{1}{2}
    \left[
      \int_{-\infty}^{\infty} \: d\tau
      \chi_{\text{eff}}(\tau)
    \right]
    \hat{n}^{2}
  }
  ,
\end{align}
where we have assumed that the pulse runs into the distant past and future for convenience.

Inserting the annihilation operator at a time ${\tau_{1}}$
now gives
\begin{align}
  &
  e^{
    -i
    \frac{1}{2}
    \left[
      \int_{\tau_{1}}^{\infty} \: d\tau
      \chi_{\text{eff}}(\tau)
    \right]
    \hat{n}^{2}
  }
  \hat{a}
  e^{
    -i
    \frac{1}{2}
    \left[
      \int_{-\infty}^{\tau_{1}} \: d\tau
      \chi_{\text{eff}}(\tau)
    \right]
    \hat{n}^{2}
  }
  \lvert \alpha \rangle
  \nonumber\\
  &\qquad=
  \left[
    \alpha
    e^{
      -i
      \frac{1}{2}
      \left[
        \int_{-\infty}^{\tau_{1}} \: d\tau
        \chi_{\text{eff}}(\tau)
      \right]
    }
  \right]
  \nonumber\\
  &\qquad\qquad\cdot
  e^{
    -i
    \left[
      \int_{-\infty}^{\tau_{1}} \: d\tau
      \chi_{\text{eff}}(\tau)
    \right]
    \hat{n}
  }
  e^{
    -i
    \frac{1}{2}
    \left[
      \int_{-\infty}^{\infty} \: d\tau
      \chi_{\text{eff}}(\tau)
    \right]
    \hat{n}^{2}
  }
  \lvert \alpha \rangle
  ,
\end{align}
and we are interested in the distribution over the phase in the operator
${ e^{ -i \phi(\tau_{1}) \hat{n} } , }$
with
\begin{align}
  \phi(\tau_{1})
  &=
  \int_{-\infty}^{\tau_{1}} \: d\tau
  \chi_{\text{eff}}(\tau)
  =
  \chi_{0} I^{2}_{0}
  \int_{-\infty}^{\tau_{1}} \: d\tau
  g^{2}(\tau)
  .
\end{align}
If ${g(\tau)}$ is Gaussian with mean ${\mu}$ and variance ${\sigma^{2}}$, then
\begin{align}
  \phi(\tau_{1})
  &=
  \frac{\chi_{0} I^{2}_{0}}{2\pi\sigma^{2}}
  \int_{-\infty}^{\tau_{1}} \: d\tau
  e^{
    -
    \left(
      \frac{\tau - \mu}{\sigma}
    \right)^{2}
  }
  ,\nonumber\\
  &=
  \frac{\chi_{0} I^{2}_{0}}{2\sqrt{\pi}\sigma}
  \Phi\left(
    \sqrt{2}
    \frac{\tau_{1} - \mu}{\sigma}
  \right)
  ,
\end{align}
where ${\Phi(\cdot)}$ is the CDF of the standard Normal distribution.

The probability of a single-photon loss event at a time ${\tau_{1}}$ is proportional to the intensity of the field at that time.
Conditioned on the occurrence of exactly one single-photon loss event across the entire pulse, the probability that this event occurs at a time ${\tau_{1}}$ is therefore governed by the normalised intensity distribution ${g(\tau_{1})}$.
Changing variables, the probability density for ${\phi}$ is
\begin{align}
  P(\phi)
  &=
  \begin{cases}
    \frac{1}{A\sqrt{2}}
    e^{
      \tfrac{1}{2}
      \left[
        \operatorname{erf}^{-1}\left( 2 \frac{\phi}{A} - 1 \right)
      \right]^{2}
    }
    ,& \phi \in (0,A) \\
    0, & \text{else},
  \end{cases}
  \nonumber\\
  A
  &=
  \frac{\chi_{0} I^{2}_{0}}{2\sqrt{\pi}\sigma}
  .
\end{align}
The mean ${\mu_{\phi}}$ for a Gaussian intensity pulse is
\begin{align}
  &
  \mu_{\phi}
  =
  \left(
    \frac{\chi_{0} I^{2}_{0}}{2\sqrt{\pi}\sigma}
  \right)
  \int^{\infty}_{-\infty} d\tau_{1}
  \Phi\left(
    \sqrt{2}
    \frac{\tau_{1} - \mu}{\sigma}
  \right)
  \frac{
    e^{
      -
      \frac{1}{2}
      \left(
        \frac{\tau_{1} - \mu}{\sigma}
      \right)^{2}
    }
  }{\sqrt{2\pi}\sigma}
  \nonumber\\
  &\quad=
  \left(
    \frac{\chi_{0} I^{2}_{0}}{2\sqrt{\pi}\sigma}
  \right)
  \int^{\infty}_{-\infty} d\tau^{\prime}_{1}
  \Phi\left(
    \sqrt{2}
    \tau^{\prime}_{1}
  \right)
  \frac{
    e^{
      -
      \frac{1}{2}
      \left(
        \tau^{\prime}_{1}
      \right)^{2}
    }
  }{\sqrt{2\pi}}
  ,\nonumber\\
  &\quad=
  \left(
    \frac{\chi_{0} I^{2}_{0}}{4\sqrt{\pi}\sigma}
  \right)
  ,
\end{align}
while for the full gate we require
\begin{align}
  \frac{2\pi}{M}
  &=
  \frac{1}{2}
  \int_{-\infty}^{\infty} \: d\tau
  \chi_{\text{eff}}(\tau)
  ,\nonumber\\
  &=
  \frac{1}{2}
  \frac{\chi_{0} I^{2}_{0}}{2\pi\sigma^{2}}
  \int_{-\infty}^{\infty} \: d\tau
  e^{
    -
    \left(
      \frac{\tau - \mu}{\sigma}
    \right)^{2}
  }
  ,\nonumber\\
  &=
  \left(
    \frac{\chi_{0} I^{2}_{0}}{4\sqrt{\pi}\sigma}
  \right)
  .
\end{align}
The mean is therefore equal to the angle of the gate, at exactly the midpoint of the range of possible values for ${\phi}$.
Rescaling ${\phi}$ by the width of this range to remove the dependence on the gate angle, we plot the rescaled distribution ${P(\phi)}$ in Figure~\ref{fig:inverted_pdf_gaussian_intensity}, compared against the uniform distribution.
This rescaled distribution is independent of the parameters ${\sigma}$ and ${\mu}$ of the Gaussian intensity pulse.

\begin{figure}[ht]
  \centering
  \includegraphics[width=0.49\textwidth]{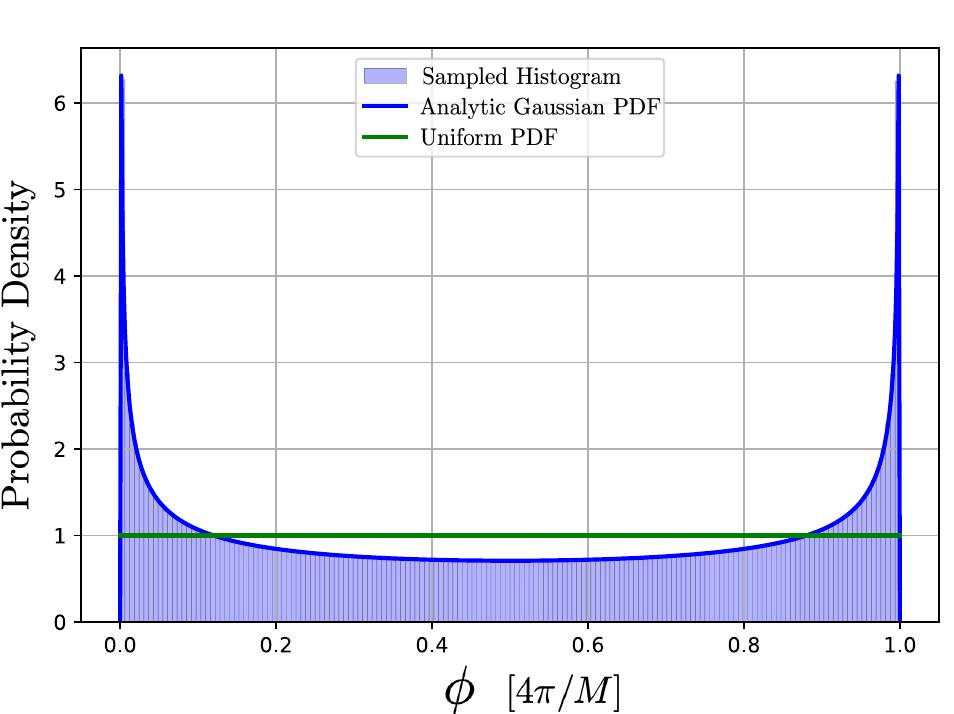}
  \caption{
    {
    The distribution over phase rotation errors conditioned on single-photon loss events, during self-Kerr evolution.
    The phase ${\phi}$ has been normalised to units of twice the total gate angle, ${4\pi/M}$, and analytic probability densities are shown for pulses of uniform (rectangular) intensity and for pulses following a Gaussian intensity profile.
    The result for the Gaussian distribution is independent of the mean time or pulse width, in these rescaled units.
    Also shown is a sampled histogram confirming the analytic shape for the Gaussian profile
    .
    }
    \label{fig:inverted_pdf_gaussian_intensity}
  }
\end{figure}

We observe from Figure~\ref{fig:inverted_pdf_gaussian_intensity} that, relative to the uniform distribution, single-photon loss events in the Gaussian case concentrate toward the end-points of the self-Kerr evolution.
The Gaussian pulse retains significant population across the full range of ${\phi}$, but this initial observation highlights the possibility that further concentration might be achieved with other pulse shapes.
If so, it would provide an approach toward retaining the protection offered by the code and the otherwise fault-tolerant circuit structure, even for loss occurring during the non-linear initialisation gates.

}

\section{Discussion}
\label{sec:discussion}

In this work, we considered the component elements of the tele-correction circuit for rotationally symmetric Bosonic codes. Broadening the class of states allowed as ancillary inputs, we observed that states generated via a generalised Yurke-Stoler procedure
\cite{leeAmplificationMulticomponentSuperpositions1994}
may be considered valid ancillary states, despite having population over the entire Fock state distribution.
{
In support, we present randomised numerical samples that demonstrate robust output state fidelities in two instances: When tele-correction subcircuits are used to modify state amplitudes, and when Yurke-Stoler states are chosen as inputs to the central, ancillary rail for tele-correction.
}

It is well documented that in using Bosonic codes we must accept a trade-off and make a balanced choice of ${\alpha}$ to effectively reduce back-action and stabilise both loss and phase errors
\cite{liCatCodesOptimal2017,albertPerformanceStructureSinglemode2018,hillmannPerformanceTeleportationBasedErrorCorrection2022}.
The main thrust of our paper has been the suitability of the Yurke-Stoler state structure for tele-correction.
However, in proposing many-component states as a bridge for non-linear interactions we have also implicitly introduced a further trade-off between the number of components, the measurement error, and the photon loss rate (which increases with ${\alpha}$).
The Yurke-Stoler states, like coherent states, contain all Fock components.
It would be interesting to see, in light of their unique phase structure, just where the balance between loss and measurement error might lie for these states.
{
An early treatment of evolution under simultaneous non-linear interaction and loss was provided in
\cite{milburnDissipativeQuantumClassical1986a}.
As our results from Section~\ref{sub:loss_during_yurke_stoler_state_initialisation} indicate however, modelling the impact of mid-evolution loss on the syndrome information obtained during measurement, to assess the circuit for fault-tolerance, will likely require a more detailed consideration and perhaps optimisation of pulse shapes.
Loss during the self-Kerr interaction used to generate the Yurke--Stoler states proves a stark example of the dangers of working only with discretised error models for Bosonic systems.
Our preliminary observations in this direction have assumed certain effects such as dispersion are negligible, and this assumption may also need to be relaxed.
}

We saw in Figure~\ref{fig:amplification_circuit_sample_results} that the modular photon number measurement could be used to turn a coherent state into a cat state, but these circuits are more broadly capable of conversion between code states of any amplitude and number of components.
{
Where non-linear interaction strength is a key limiting factor, we propose using many-component Yurke-Stoler states as mediating ancillae to shorten interactions times for such inter-conversion, facilitating ancillary state generation and repeated correction.
}

{
Finally, although the use of Yurke--Stoler states as proposed here allows loss to be traded against the non-linear interaction strength, the requirements for implementing such non-linearities in optical systems remain highly demanding.
Our brief survey of non-linear materials in Appendix~\ref{sec:utility_thresholds_for_the_third_order_nonlinear_susceptibility} suggests looking to near-resonant effects in coherent ensembles.
Recent work has demonstrated the storage and retrieval of single photons in a hot rubidium vapour ensemble quantum memory
\cite{thomasDeterministicStorageRetrieval2023}.
It would be interesting to explore if such a system could be adapted to generate coherent nonlinearities in stronger pulses.
}

.

\begin{acknowledgments}
    This work was supported by the UK EPSRC through EP/Y004752/1, EP/W032643/1, EP/Z53318X/1 and the National Research Foundation of Korea (NRF) grant funded by the Korea government (MSIT) (No. RS-2024-00413957).
    NL, SN, WJM and KN were supported in part from the Moonshot R\&D Program Grants JPMJMS2061 \& JPMJMS226C, COI-NEXT under Grant No. JPMJPF2221, and the JSPS KAKENHI Grant No. 21H04880.
    SN further acknowledges support from JSPS KAKENHI Grant Number JP22KJ1436.
\end{acknowledgments}

The data that support the findings of this article are openly available at~\cite{hanksarticledataset2026}.

\appendix

{

\section{What is Meant by Fault-Tolerance}
\label{sec:what_is_meant_by_fault_tolerance}

Here we discuss the meaning of fault tolerance in the context of single-mode bosonic codes, which are approximate quantum error correction codes.

Fault-tolerance can have slightly different meanings depending on context and implication.
When applied generally, fault-tolerance can refer to the existence, for a given error model, of an error correction threshold. For physical error rates below this threshold, the implication is then that logical errors can be suppressed arbitrarily with an overhead in resources scaling logarithmically with respect to the desired reduction in the error rate (e.g. an exponential suppression for a polynomial overhead).
Crucially, and distinguishing fault tolerant quantum error correction from the information-theoretic task of efficient code design, the error model must include the possibility of errors occuring within the error correction circuitry itself
\cite{shorFaulttolerantQuantumComputation1996}.
To leading order, a failure in such circuitry should have no worse an effect than a failure in any other unreliable element.
This sense of fault-tolerance as the ability to scale the code arbitrarily to suppress the logical error rate cannot however apply to single-mode bosonic codes, because there is a necessary trade-off in error suppression between conjugate variables
\cite{liCatCodesOptimal2017,ouyangTradeOffsNumberPhase2021}
(e.g. position and momentum, or number and phase).

The concept of fault-tolerance can nonetheless be useful even for the bounded and approximate error correction of single-mode bosonic codes, when used in its restricted sense as applied to the structures of quantum circuits.
Loosely, when interactions in a circuit can propagate errors, a fault-tolerant implementation will preserve the dependence (to leading order) of the logical error rate on the code distance --- small physical faults (loss, dephasing, control/ancilla errors) should not cascade into uncorrectable logical errors.
For discrete, multi-qubit codes with local noise, this means preventing the spread of errors between physical qubits within a single code block.

Code distances for the single-mode bosonic codes we consider are defined for phase errors as rotation angles in phase space and for loss errors as the separation in photon number between populated elements in the Fock basis.
For phase rotation errors, we have shown in Appendix~\ref{sec:measurement_error_through_total_variation_distance} by approximation of the total-variation distance that there is an exponential suppression in the error of logical measurement as the coherent state components become widely separated.
This result can be directly applied to the phase rotation error model, for the evolution operator
${\hat{U}_{\tau} = e^{i\theta {\tau} \hat{a}^{\dagger}\hat{a}}}$,
by choosing a time step size ${\tau}$ to define the increment ${\theta \tau}$ as a \emph{characteristic} increment of local phase error associated with a gate, subroutine or other \emph{unit}.
The logical error rate per unit then scales exponentially in ratio between the angular code distance and the magnitude of this small increment.
For the loss error model, a similar approach can be taken in the super-operator picture. On vectorising the density matrix, the pure loss channel takes the Gaussian form
${(1-\gamma)^{ \frac{1}{2} \left( \hat{a}^{\dagger}\hat{a} + \hat{b}^{\dagger}\hat{b} \right) } e^{ \gamma \hat{a} \hat{b} }}$,
where ${\hat{b}}$ is the transpose of the creation operator applied from the right in the normal density matrix picture.
At small times, the right-most factor can be linearised as
${1 + \gamma_{\tau} \hat{a} \hat{b}}$,
with `local' error in this case corresponding to a single-photon loss event.
The probability of logical error corresponding to a shift in the Fock basis, for code distance ${K}$, scales to leading order in the exponential as ${(\gamma_{\tau} \lvert\alpha\rvert^{2})^{K}}$.
The left-most factor, scaling coherent state components as
${\lvert \alpha \rangle \rightarrow \lvert \alpha \sqrt{1-\gamma_{\tau}}\rangle}$,
has the effect of reducing the inter-component spacing (and thus the effective angular code distance) by
${\alpha \gamma_{\tau} \sin(\pi/K)}$ in the small-${\gamma}$ limit (by Taylor series expansion).

A fault-tolerant circuit should preserve the magnitude of logical errors to leading order in ${\theta {\tau}}$ and in ${\gamma_{\tau}}$.
The tele-correction circuit makes extensive use of the rotation gate ${\hat{R}_{MK}}$. This gate commutes with phase rotation errors, but transforms photon loss events on either input subsystem into phase rotation errors in the other.
Specifically,
\begin{align}
   \hat{R}_{MK} \hat{a}_{1}
    &=
    \hat{a}_{1}
    e^{-i\frac{4\pi}{MK} \hat{n}_{2}}
    \hat{R}_{MK}
    .
\end{align}
There is a manifestation here of the trade-off between the protection offered by single-mode bosonic codes against number and phase error: increasing the numbers ${M}$ and ${K}$ decreases the angular code distances, but to preserve the magnitude of phase errors to leading order we nonetheless require a choice of ${M}$ and ${K}$ so that the angle ${\frac{4\pi}{MK}}$ is consistent with the larger of ${\gamma_{\tau}}$ and ${\theta_{\tau}}$.
In this way, the demand for fault-tolerance in the telecorrection circuit puts an upper bound on the code distance, once characteristic error magnitudes have been defined.

Aside from the propagation of existing errors, additional sources of noise arising in the circuit itself are the initialisation noise in newly introduced ancillae, readout back-action, and control imperfections.
While discretised error models are common in the analysis of logical error in quantum error correction, realistic noise for bosonic codes may depart significantly from this picture (we discuss an example of this in Section~\ref{sub:loss_during_yurke_stoler_state_initialisation}).

}

\section{Measurement Error through the Total Variation Distance}
\label{sec:measurement_error_through_total_variation_distance}

We can look at the probability of error in the final state to leading order by concentrating only on incorrect measurement outcomes with support on neighbouring coherent state components.
The total variation distance (TVD) and associated single trial error ${p_{\text{err}}}$ for distinguishing between two Gaussian distributions of equal variance are given by
\begin{align}
  \mathrm{TVD}
  &=
  \mathrm{erf}
  \left(
    \frac{\lvert\mu_{1}-\mu_{2}\rvert}{2\sqrt{2}\sigma}
  \right)
  ,\nonumber\\
  p_{\text{err}}
  &=
  \frac{1 - \mathrm{TVD}}{2}
  .
\end{align}
When the coherent state components are distributed evenly in phase space at ${2D}$ points around a circle of radius ${\lvert\alpha\rvert}$,
the distance between two components $\vert {\alpha} \rangle$ and $\vert {\alpha} e^{\frac{i\pi}{{D}}} \rangle$ is ${2 \lvert\alpha\rvert \sin\left(\frac{\pi}{2D}\right)}$, as shown in Figure~\ref{fig:cat_project_distance_sketch}.
The total variation distance then becomes
\begin{align}
  \mathrm{TVD}_{\alpha,D}
  &=
  \mathrm{erf}
  \left(
    \sqrt{2} \lvert\alpha\rvert \sin\left(\frac{\pi}{2D}\right)
  \right)
  ,\nonumber\\
  &\approx
  1 - e^{-\frac{1}{2}\left( 2 \lvert\alpha\rvert \sin\left(\frac{\pi}{2D}\right)  \right)^{2}}
  ,
\end{align}
where the final line approximates the error function to only the first order term in
the B\"{u}rmann expansion
\cite{schopfBurmannsTheoremIts2014}.
When the local oscillator of the heterodyne measurement is weak, the measurement error depends on the discrete lattice of heterodyne detection outcomes and must be determined numerically.

{
\section{Utility Thresholds for the Third-order Nonlinear Susceptibility}
\label{sec:utility_thresholds_for_the_third_order_nonlinear_susceptibility}

The Hamiltonian terms in a perturbative expansion involving electric fields at fourth order are proportional to
\cite{gerryIntroductoryQuantumOptics2004,boydNonlinearOptics2019}
\begin{align}
  V
  \varepsilon_{0}
  \chi^{(3)}
  \mathcal{E}^{4}_{0}
  ,\quad
  \mathcal{E}_{0}
  =
  \sqrt{
      \frac{
        \hbar \omega
      }{
        V \varepsilon_{0}
      }
  }
  ,
\end{align}
where ${V}$ is the volume populated by the field, ${\omega}$ is its angular frequency, and ${\chi^{(3)}}$ is the third-order non-linear susceptibility of the medium.
We have suppressed any dependence of ${\chi^{(3)}}$ on frequency or polarisation, and are treating the field distribution as an average over the volume ${V}$.

To generate the ${N}$-component Yurke--Stoler states, we require
\begin{align}
  \frac{V}{\hbar}
  \varepsilon_{0}
  \chi^{(3)}
  \mathcal{E}^{4}_{0}
  \tau
  &\sim
  \frac{\pi}{N}
  ,
\end{align}
with ${\tau}$ the interaction time (for which, again, an averaged interaction strength has been assumed).
The physical constants are approximately
\begin{align}
  \hbar
  &\approx
  10^{-34}
  \mathrm{\:J\cdot s}
  ,\nonumber\\
  \varepsilon_{0}
  &\approx
  (36\pi)^{-1} \times 10^{-9}
  \mathrm{\:F\cdot m^{-1}}
  ,\nonumber\\
  c
  &\approx
  3\times 10^{8}
  \mathrm{\:m\cdot s^{-1}}
  .
\end{align}
For fields in the optical--telecom range we have
${\omega \sim 4\pi \times 10^{14} \mathrm{\:rad\:s^{-1}}}$,
and as a rough approximation we take a beam cross-section at the diffraction limit, of scale
${\left(\frac{\pi c}{\omega}\right)^{2}\approx 10^{-12} \mathrm{\:m^{2}}}$
(${\sim1\mathrm{\:\mu m^{2}}}$ \cite{hosseiniAnalysisKerrNonlinearity2018}),
with a characteristic pulse time ${\tau_{p}}$ resulting in an effective beam length ${c \tau_{p}}$.
Substituting these quantities into the equation above gives
\begin{align}
  N \chi^{(3)}
  &\sim
  V \tau^{-1}
  \frac{
    \pi \varepsilon_{0}
  }{
    \hbar \omega^{2}
  }
  ,\nonumber\\
  &\approx
  \frac{\tau_{p} \tau^{-1}}{192\pi^{2}}
  10^{-7}
  .
\end{align}
The characteristic pulse time ${\tau_{p}}$ might be anywhere in the range ${10^{-8}\leftrightarrow 10^{-12} \mathrm{\:s}}$ depending on laser characteristics and timing accuracy constraints, where the lower end of this range corresponds to a
${\sim 1\%}$
relative width for a central angular frequency ${\omega}$.
The interaction time ${\tau}$ is limited by the attenuation rate in the medium, which for many optical nonlinear media is expressed on a ${\mathrm{cm}}$ scale
\cite{nikogosyanBasicNonlinearOptical2005,tuCharacterizationOptimalDesign2020},
though there is wide variation.
As a first approximation we take the interaction length to be on the ${0.1\mathrm{\:mm}}$ scale, which for a typical range of condensed matter refractive indices ${1 \leq n \leq 3}$ \cite{boydNonlinearOptics2019}, suggests an interaction time ${\tau}$ in the range
${10^{-12} \leftrightarrow 10^{-13} \mathrm{\:s}}$.
We then have a range of estimated threshold values for ${N \chi^{(3)}}$,
\begin{align}
  5 \times 10^{-11}
  &\lesssim
  N \chi^{(3)}
  \lesssim
  5 \times 10^{-6}
  .
\end{align}
At and above this range, the generation of ${N}$-component Yurke--Stoler states might be feasible in principle.
}

{

\subsection{Surveying Nonlinear Materials}
\label{sub:nonlinear_materials_strengths_and_weaknesses}

The nonlinear susceptibilities of many materials are too small for the application described in this work.
For example, solid materials described in Boyd \cite{boydNonlinearOptics2019} with the largest ${\chi^{(3)}}$ values are:
\begin{center}
  \begin{tabular}{r|l}
    Material & ${\chi^{(3)} \: (\mathrm{m^{2}V^{-2}})}$ \\
    \hline
    GaAs & ${1.4\times 10^{-18} }$ \\
    Si & ${2.8\times 10^{-18} }$ \\
    As${_{2}}$S${_{3}}$ glass & ${4.1\times 10^{-19} }$ \\
    CS 3-68 nanoparticles in glass & ${1.8\times 10^{-16} }$ \\
    gold nanoparticles in glass & ${2.1\times 10^{-16} }$ \\
    PTS polymer & ${-5.6\times 10^{-16} }$ \\
    4BCMU polymer & ${-1.3\times 10^{-19} }$ \\
    Ag & ${2.8\times 10^{-19} }$ \\
    Au & ${7.6\times 10^{-17} }$ \\
  \end{tabular}
\end{center}
Nonlinearities in silicon--nitride waveguides at ${\sim1.5\mathrm{\:\mu m}}$ have seen much development in the past several years \cite{ooiPushingLimitsCMOS2017,hosseiniAnalysisKerrNonlinearity2018,tu2MidinfraredSiliconrich2021,permanaInvestigationKerrNonlinearity2024}.
As representative numbers,
Krückel et al.
\cite{kruckelOpticalBandgapEngineering2017} measure
\begin{align}
  n_{2} &= 1.1\times 10^{-18}\mathrm{\:m^{2}W^{-1}}
  ,\nonumber\\
  n_{0,\mathrm{eff}} &= 2.2
  ,\quad
  ( \chi^{(3)} = 1.9 \times 10^{-20} )
  ,
\end{align}
while
Lacava et al. \cite{lacavaSirichSiliconNitride2017}
observe
\begin{align}
  n_{2} &= 2 \times 10^{-18}\mathrm{\:m^{2}W^{-1}}
  ,\nonumber\\
  n_{0,\mathrm{eff}} &= 2.71
  ,\quad
  ( \chi^{(3)} = 5.2 \times 10^{-20} )
  .
\end{align}
Alam et al. \cite{alamLargeOpticalNonlinearity2016} report ${n_{2}}$ in indium tin oxide at ${1240\mathrm{\:nm}}$ of
\begin{align}
  n_{2,\text{(eff)}}
  &=
  0.11
  \mathrm{\:cm^{2}/GW}
  ,\nonumber\\
  &=
  1.1 \times
  10^{-14}
  \mathrm{\:m^{2}W^{-1}}
  \quad
  (
    \chi^{(3)}
    \approx
    1.3 \times 10^{-16}
  )
\end{align}
via laser-induced heating of electrons in the conduction band
(taking ${n_{0}=1.8}$ \cite{polyanskiyRefractiveindexinfoDatabaseOptical2024}).
We have used the relation
\cite{boydNonlinearOptics2019}
\begin{align}
  n_{2}
  &=
  \frac{
    3
  }{
    4 n_{0} \mathrm{Re}[n_{0}] \varepsilon_{0} c
  }
  \chi^{(3)}
\end{align}
in the formulae above.
The susceptibilities of all these materials are several orders of magnitude below our lower threshold ${\sim10^{-11}}$, so it is unlikely that they can be used to implement our proposal even for very large ${N\sim 100}$.

Sychev et al. \cite{sychevAllopticalModulationSingle2025} have recently reported very high non-linear refractive indices in a silicon SPAD, of
\begin{align}
  n_{2}
  &=
  -3.3 \times 10^{-4}
  \mathrm{\:m^{2}W^{-1}}
  \quad
  (
    \chi^{(3)}
    =
    -1.4 \times 10^{-5}
  )
  ,
\end{align}
and
\begin{align}
  n_{2}
  &=
  1.3 \times 10^{-2}
  \mathrm{\:m^{2}W^{-1}}
  \quad
  (
    \chi^{(3)}
    =
    5.6 \times 10^{-4}
  )
  ,
\end{align}
(taking ${n_{0}=3.5}$ \cite{polyanskiyRefractiveindexinfoDatabaseOptical2024})
arising respectively from an electron avalanche effect populating the conduction band, and subsequent thermalisation.

However, rapidly thermalising processes may not be suitable in applications where it is necessary to maintain quantum coherence.
If ${\hat{E}_{l}}$ represents the Kraus operator for ${l}$-photon loss
\cite{albertPerformanceStructureSinglemode2018},
then, operating in a macroscopic medium well below any saturation limit,
an intensity-dependent phase shift induced via thermal effects (or other absorbing processes creating rapidly-decohering macroscopic states)
will likely have a form similar to
\begin{align}
  \sum^{\infty}_{l=0}
  e^{i \theta_{l} \hat{n}}
  \hat{E}_{l}
  \hat{\rho}
  \hat{E}^{\dagger}_{l}
  e^{-i \theta_{l} \hat{n}}
  ,\quad
  \theta_{l}
  \propto
  l
  .
\end{align}
A channel of this kind will produce an intensity-dependent phase shift, as the probability of an ${l}$-photon loss event depends explicitly on the mean photon number of the pulse. However, it will not do so via the coherent self- or cross-Kerr evolution assumed in this manuscript, and cannot create the entanglement necessary for telecorrection.

Caspani et al. \cite{caspaniEnhancedNonlinearRefractive2016} exploit a related mechanism to that of Alam et al. mentioned above, but in aluminium-doped zinc oxide, achieving
${\lvert \chi^{(3)} \rvert \sim 10^{-19}}$ at ${1.5\mathrm{\:\mu m}}$.
The novel aspect of their work that may be relevant to our purposes is the use of a strong pump beam to modify the effective linear refractive index. This allows them to modify the values ${n_{0} \mathrm{Re}[n_{0}]}$ we used in the relation above, increasing the sensitivity of the nonlinearity while avoiding loss in the weaker probe beam.
Though we are led to observe then that rapidly thermalising processes are likely unsuitable, applied directly, it could be interesting to explore whether coherent nonlinearities could be similarly enhanced by thermalising pumps.

While thermalising absorption processes may lose coherence, virtual processes far below-resonance have the frequency-independent form
\cite{boydNonlinearOptics2019}
\begin{align}
  \chi^{(3)}
  &\approx
  \frac{
    8 N \mu^{4}
  }{
    \varepsilon_{0}
    \hbar^{3}
    \omega^{3}_{0}
  }
  ,
\end{align}
where ${\mu}$ is a typical dipole matrix element, ${N}$ a molecular density, ${\omega_{0}}$ an atomic resonance frequency, and a typical value is of order
${10^{-22}\mathrm{\:m^{2}V^{-2}}}$.
Closer to resonance, strong effective nonlinearities have been implemented with low-dimensional systems
\cite{bruneManipulationPhotonsCavity1992,liuPreparationMacroscopicQuantum2005},
and measurement-based schemes have been proposed based on single-photon detection \cite{scheelMeasurementinducedNonlinearityLinear2003,costanzoMeasurementInducedStrongKerr2017,chalermpusitarakProgrammableGenerationArbitrary2025}. These approaches do not preserve the fault-tolerant property of the telecorrection circuit, however.

Turning to coherent ensembles, in an ultracold atomic gas in 1999 Hau et al.
\cite{hauLightSpeedReduction1999}
measured
\begin{align}
  n_{2,\text{(eff)}}
  &=
  0.18
  \mathrm{\:cm^{2}/W}
  ,\nonumber\\
  &=
  1.8 \times
  10^{-5}
  \mathrm{\:m^{2}W^{-1}}
  \quad
  (
    \chi^{(3)}
    \approx
    6.4
    \times 10^{-8}
  )
\end{align}
(which we have assumed is dilute).
Such ensembles seem promising.

}

{

\section{Random Displacement Width from Relative Intensity Noise}
\label{sec:random_displacement_width_from_relative_intensity_noise}

To assess whether the chosen scaling $\sigma^\prime$ of the coherent state displacement in Section~\ref{sub:displacement_noise_at_the_input} is physically reasonable, it is useful to relate it to experimentally measurable quantities such as the intensity noise (optical power fluctuations) of a laser. The relative intensity noise (RIN) is the power noise normalised to the average power level
\cite{Paschotta_2007_relative_intensity_noise}.
Then the root mean square (rms) value of the power for a RIN distribution, constant over a given frequency range $\Delta \omega$, is
\begin{equation}
    \frac{\Delta P}{\bar{P}} |_{rms} \sim {RIN \cdot \Delta \omega},
\end{equation}
when RIN is quoted in units $\mathrm{Hz}^{-1}$. This expression corresponds to the standard deviation in the mean photon number per unit time, integrated across the relevant frequency range, under the assumption that the frequency distribution is flat in the region of interest. We now consider a typical RIN measurement value $\sim 10^{-15}\,\mathrm{Hz}^{-1}$
and the frequency range $\Delta\omega \sim 10^9 $Hz based on a typical laser pulse length $\sim 1$ns
\cite{aghaeeradScalingNetworkingModular2025}.
We then deduce an estimate: 
\begin{equation}
\begin{aligned}
    \frac{\Delta P}{\bar{P}} |_{rms} &\sim 10^{-6} \\
    &\sim \frac{K}{2}.
\end{aligned}
\end{equation}
with spread factor $K = \Delta \omega / \omega$ where $\omega$ is the characteristic optical frequency --- visible lasers are in the region of hundreds of terahertz, i.e. $\omega \sim 5 \times 10^{14}$~Hz.

Expanding the rms relative power fluctuations in terms of mean photon number, we expect
\begin{equation}
    \left(\frac{\Delta P}{\bar{P}} |_{rms} \right)^2 = \frac{\langle\hat{a}^{\dagger2} \hat{a}^2\rangle + \langle\hat{a}^\dagger \hat{a}\rangle -\langle\hat{a}^\dagger \hat{a}\rangle ^2}{\langle\hat{a}^\dagger \hat{a}\rangle^2} - \frac{1}{|\alpha|^2} 
\end{equation}
where the term $1/|\alpha|^2$ is further subtracted to take into account the variance of the input coherent state (shot noise limit). For our noisy state of interest, we identify the coherent state input as a displaced thermal state, where the initial thermal state prior to displacement has mean photon number $\bar{n} = 2\sigma^2$. Using the known formula for general normally-ordered moments of the displaced thermal state
\cite{cahillDensityOperatorsQuasiprobability1969, marianSqueezedStatesThermal1993},
\begin{equation}
    \langle \hat{a}^{\dagger q} \hat{a} ^{r} \rangle = q! \alpha ^{r-q} \bar{n}^q L_q^{(r-q)} (-|\alpha|^2 /\hat{n}), \quad r \geq q,
\end{equation}
we can identify the mean photon number after displacement as $\bar{n} + |\alpha|^2$.
\begin{equation}
\begin{aligned}
    \frac{\Delta P}{\bar{P}} |_{rms}  & = \sqrt{\frac{\frac{1}{N} \sum_i (P_i - \bar{P})^2}{\bar{P}}}\\
    &= \sqrt{\frac{\langle (\hat{a}^{\dagger} \hat{a} - \langle \hat{a}^{\dagger} \hat{a}\rangle )^2\rangle}{\langle \hat{a}^{\dagger} \hat{a}\rangle^2}}
\end{aligned}
\end{equation}
With normal-ordering and squaring both sides we retrieve the earlier equation.
The Laguerre polynomial ${L_2^{(0)} (x) = (x^2 -4x + 2)/2}$
\cite{abramowitzHandbookMathematicalFunctions1972}
gives
\begin{equation}
\begin{aligned}
        \langle\hat{a}^{\dagger2} \hat{a}^2\rangle &= 2! \bar{n}^2 L_2^{(0)} (-|\alpha|^2
/\bar{n})\\
&= |\alpha|^4 + 4|\alpha|^2 \bar{n} + 2 \bar{n}^2,
\end{aligned}
\end{equation}
leading to the relative power fluctuations in terms of $\bar{n}$ and $\alpha$: 
\begin{equation}
\begin{aligned}
    \left(
      \frac{\Delta P}{\bar{P}} |_{rms}
      \right)^2
    &=
    \frac{
      (|\alpha|^4 + 4|\alpha|^2 \bar{n} + 2 \bar{n}^2) + (\bar{n} + |\alpha|^2)
    }{
      (\bar{n} + |\alpha|^2)^2
    }
    -
    1
    - \frac{1}{|\alpha|^2}
    \\
    &=
    \frac{
      2|\alpha|^2 \bar{n}+ \bar{n}^2 + \bar{n} + |\alpha|^2
    }{
      \bar{n}^2 + 2\bar{n}|\alpha|^2 + |\alpha|^4
    }
    - \frac{1}{|\alpha|^2}
    \\
    &\sim
    (10^{-6})^2.
\end{aligned}
\end{equation}

}

\end{document}